\documentclass[%
 reprint,
 amsmath,amssymb,
 aps,
]{revtex4-2}

\usepackage{graphicx}
\usepackage{dcolumn}
\usepackage{bm}

\begin{document}

\preprint{APS/123-QED}

\title{Hybrid Coupling Topology with Dynamic ZZ Suppression for Optimizing Circuit Depth during Runtime in Superconducting Quantum Processor}

\author{Uday Sannigrahi}
\email{udaysannigrahi@gmail.com}
\affiliation{University of Calcutta, Kolkata, India}
\altaffiliation{Also at University of Calcutta, Kolkata, India}

\author{Amlan Chakrabarti}
\email{acakcs@caluniv.ac.in}
\affiliation{University of Calcutta, Kolkata, India}

\author{Swapnil Saha}
\affiliation{Jadavpur University, Kolkata, India}
\altaffiliation{Also at Jadavpur University, Kolkata, India}

\author{Shrinjita Biswas}
\affiliation{Jadavpur University, Kolkata, India}

\date{\today}

\begin{abstract}
To reduce circuit depth when executing Quantum algorithms, it is necessary to maximize qubit connectivity on a near-term quantum processor. While addressing this, we also need to ensure high gate fidelity, suppression of unwanted ZZ cross-talk, a compact layout footprint, and minimal control hardware complexity to support scalability. In current superconducting quantum chips, fixed coupling is used as it is easier to scale, but it is limited by unwanted static ZZ interaction during single qubit operations, which degrades system performance. To overcome these challenges, we have introduced a first-of-its-kind hybrid tunable-coupling architecture that connects four fixed-frequency transmon qubits using a single coupler. This hybrid coupler uses off-resonant Stark drives to tune ZZ strength between qubit pairs. Experimentally backed simulation results indicate that our proposed hybrid design maximizes the qubit connectivity while reducing control overhead. This design achieves a near 20\% reduction in circuit depth compared to IBM's Heavy-Hexagonal layout, showing its potential for scalability. 
\end{abstract}

\maketitle


\section{Introduction}
{S}{uperconducting} qubit's \cite{Nakamura1999CoherentCO} primary figure of merit has been dominated by relaxation time ($T_1$) and dephasing time ($T_2$). These $T_1$ and $T_2$ are dominated by dielectric losses in the material and the purity of the fabricated design. From the reported works, this problem is being progressed towards near-millisecond energy relaxation and dephasing times for a superconducting transmon qubit \cite{REF1}. They have achieved (666 ± 33)us of $T_1$  and (1057 ± 138)us
of $T_{2echo}$.To maximize computational power in a superconducting quantum computer, which can outperform a classical computer, high-fidelity, fast entangling gates are required to ensure that gate lengths are orders of magnitude shorter than $T_1$ and $T_2$. \par

Achieving high fidelity gates, gate speed is one of the dominating factors. Single qubit gate speeds are limited by anharmonicity($\eta$) of the transmon qubit, which is near~200-300 MHz range. $\eta$ is highly dependent on the fabrication of the transmon qubit.\par
High-fidelity and high-speed two-qubit entangling gates are a fundamental need for executing algorithms that surpass classical computational capabilities. To achieve this, we need high interaction strength between the qubit pair to maximize gate speed, suppress ZZ crosstalk during the idle state, and advance pulse shaping techniques to suppress leakage into non-computational states.\par
To solve this, some of the works reported on tunable coupling architecture on superconducting qubit technology. \cite{PhysRevApplied.10.054062} proposes a flux-tunable coupler to perform a two-qubit gate. Another work \cite{Geller2014TunableCF} reported a high-performance tunable coupler suitable for superconducting Xmon and planar transmon qubits, but still limited by qubit-qubit connectivity. A high-fidelity two-qubit gate can also be realized with microwave-activated \cite{PhysRevLett.127.200502} tunable coupling, where the ZZ interaction strength can be tuned by applying off-resonant drive pulses simultaneously. \par

IBM's Eagle \cite{ibm2021eagle} and Google's Sycamore \cite{Arute2019} use the well-known transmon-coupler-transmon structure, where the coupler connects two transmon qubits. Using this topology, they have enhanced the hardware performance through error-mitigation techniques, but the hardware limits the number of connections between qubits. \par

For a practical QPU, one of the key challenges is to achieve maximum qubit-qubit coupling pairs for implementing two-qubit gates and reducing the circuit depth by inserting fewer SWAP gates while executing quantum algorithms. For an optimal layout footprint with the maximum number of coupled qubits, we should move our focus to sophisticated coupling architecture design from the widely adopted qubit-coupler-qubit \cite{PhysRevLett.113.220502} architecture. \par

Although IBM's heavy hexagonal layout offers better stability and high gate fidelity through error mitigation and quantum error corrections, it increases the circuit depth drastically by inserting many SWAP gates \cite{Li2023}, which hampers the algorithmic fidelity during execution of the quantum algorithm. In recent years, the industry has been moving towards tunable coupler transmon architecture, but it still requires more layout footprint by introducing another flux line for each pair of coupled transmon, which also increases the control line complexity. \par

Motivated by these challenges and the growth of new coupling techniques to address the high-fidelity bottleneck with maximum connectivity, we have proposed a design that increases qubit connectivity while deploying only the maximum number of fixed-frequency transmons. We have used a flux-tunable transmon in each cluster, with an off-resonant Stark-drive scheme, to enable strong, tunable ZZ interactions.
The key contributions of this work can be stated as follows:
\begin{itemize}
    \item Our proposed design enhances qubit isolation and enables strong tunable ZZ interactions for implementing fast entangle gates.
    \item This design achieved enhanced ZZ tunability while requiring only a single flux control line per qubit cluster. The design reduces the complexity of the control hardware and improves scalability.
    \item We demonstrate a reconfigurable coupling topology, and we have benchmarked it against IBM's backend Heron processor. Our proposed design achieves a near 20\% reduction in overall circuit depth, demonstrating its practical benefits for near-term quantum computation.
\end{itemize}

The remainder of the paper is organized as follows. Section II provides a brief overview of coupling architectures. Section III presents the methodology of the proposed design. Section IV discusses the analysis of the results. Section V concludes the paper.

\section{Related research work}

Early coupling architecture designs were primarily based on fixed capacitive coupling architectures \cite{PhysRevA.76.042319}. These designs implemented microwave-drive schemes for universal two-qubit gates that work by driving one qubit(control) at the transition frequency of the target qubit\cite{PhysRevB.81.134507}. As qubit counts increase, one of the major drawbacks of these designs is frequency crowding and slow gate times, which put a limitation on achievable circuit depth. \par

In recent years, IBM and other major players have incorporated tunable couplers in their designs. Tunable couplers mitigate the always-on ZZ interactions and offer a faster gate. The coupling strength is modulated via magnetic flux \cite{PhysRevApplied.10.054062}. These designs offer high on/off ratios by destructively interfering with the direct coupling pathway. However, the routing of flux lines increases signal crosstalk and makes the system exposed to 1/f noise, which results in faster decoherence \cite{PhysRevApplied.12.054023}.

The physical layout of these tunable couplers dictates the circuit depth of quantum circuit compilation. IBM's heavy-hex lattice limits the number of coupling pairs, which has a direct impact on circuit depth. Low connectivity leads to more SWAP operations while executing quantum algorithms. To address this to some extent, one of the recent works demonstrated a novel architecture where they have used a single fixed-frequency resonator coupler to couple three transmon qubits \cite{Kang2024NewDO}. This reduced the hardware overhead while increasing the coupling connectivity. \par

Despite this advancement, the circuit depth optimization can  be possible by coupling all qubits to each other in a grid format while ensuring suppressed ZZ interaction and compact layout footprints. Our hybrid design optimizes the qubit-qubit connectivity  with $ZZ$ tunability, and the coupling map can be modified during runtime for optimum circuit depth.

\begin{figure}
    \centering
    \includegraphics[width=0.5\linewidth]{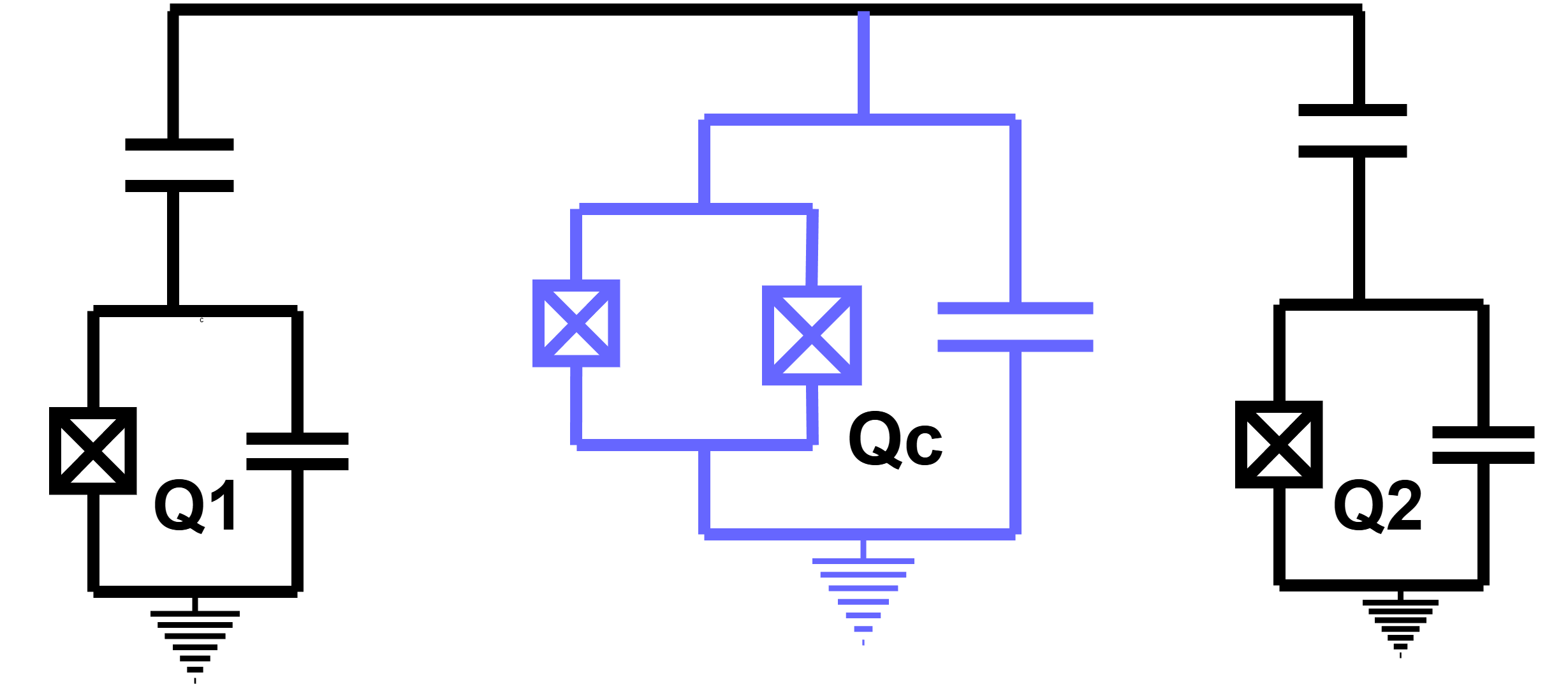}
    \caption{Circuit schematic of the $Q_1-Q_c-Q_2$ coupling architecture. The diagram illustrates the fundamental building block of the proposed processor, consisting of two computational transmon qubits ($Q_1$ and $Q_2$) capacitively coupled to a central flux-tunable transmon coupler ($Q_c$).}
    \label{fig:three_transmon_coupling}
\end{figure}

\section{Methodology}
For the analysis of the coupling architecture we consider a linear chain of three transmons as depicted in Fig \ref{fig:three_transmon_coupling},
where $Q_{1}$ and $Q_{2}$ are fixed frequency transmon with parameters( $\omega_{1}/2\pi =5.24GHz$), $\eta_{1}/2\pi = -0.215 MHz$,( $\omega_{1}/2\pi =5.02GHz$), $\eta_{1}/2\pi = -0.209 MHz$, coupled by a flux-tunnable transmon $Q_{c}$. In the drive  frame taking Duffing oscillator approximation of the transmon, the Hamiltonian is given by:
\begin{equation}
\begin{aligned}
H =\;&
\sum_{i \in \{1,2\}}
\left[
(\omega_i - \omega_d)\hat{a}_i^\dagger \hat{a}_i
+ \frac{\eta_i}{2}\hat{a}_i^{\dagger}\hat{a}_i^{\dagger}\hat{a}_i\hat{a}_i
\right] \\
&+
\left[
(\omega_c(\Phi )\hat{a}_c^\dagger \hat{a}_c
+ \frac{\eta_c}{2}\hat{a}_c^{\dagger}\hat{a}_c^{\dagger}\hat{a}_c\hat{a}_c
\right] \\
&+ \sum_{k=1,2} J_{kc}
\left(\hat{a}_k^\dagger \hat{a}_c + \hat{a}_c^\dagger \hat{a}_k\right) \\
&+ \sum_{i \in \{1,2\}}
\left(\epsilon_i \hat{a}_i + \epsilon_i^{*}\hat{a}_i^\dagger\right)
\end{aligned}
\end{equation}
where $\omega_i /2\pi$ is the transition frequency between $|0>$ and $|1>$ for \textit{$i_{th} $}transmon  , the drive frequency (off-resonant) is $\omega_d$, $a_i$ and $a_i^{\dagger}$ are bosonic annhilation and creation operator, $\eta_i$ is the anharmonicity, $J_{kc}$ is the coupling strength between transmon $k$ and the central coupler transmon $c$ and $\epsilon_i$ is the complex drive amplitude.
The transition frequency $\omega_c{(\Phi)}$ of the coupler $Q_c$ is tuned by using an external flux $\Phi(t)$. The detunning $\Delta_{ic}$ between $Qc$ and $Q_{i}$ ($i=1,2$) is negative. \par

 The combined system is modeled as a three-transmon Hamiltonian as shown in Eq. 1
We choose   $\Phi$ such that the $\omega_c$ is detuned by several $>200 MHz$ during suppression of ZZ. For strong two-qubit interaction, we bring $Qc$ into a regime $\approx200MHz$ away from $Q_{1,2}$. We apply two off-resonant drives to Q1 and Q2, following the Stark-shift scheme \cite{PhysRevLett.127.200502}. Driving with an off-resonant EM field induces state-dependent AC Stark shifts. The effective $\zeta$ can be modified by increasing drive amplitude. Choosing a relative phase even cancels $\zeta$, which is shown in \cite{PhysRevLett.127.200502} experimentally.

\begin{figure}
    \centering
    \includegraphics[width=1\linewidth]{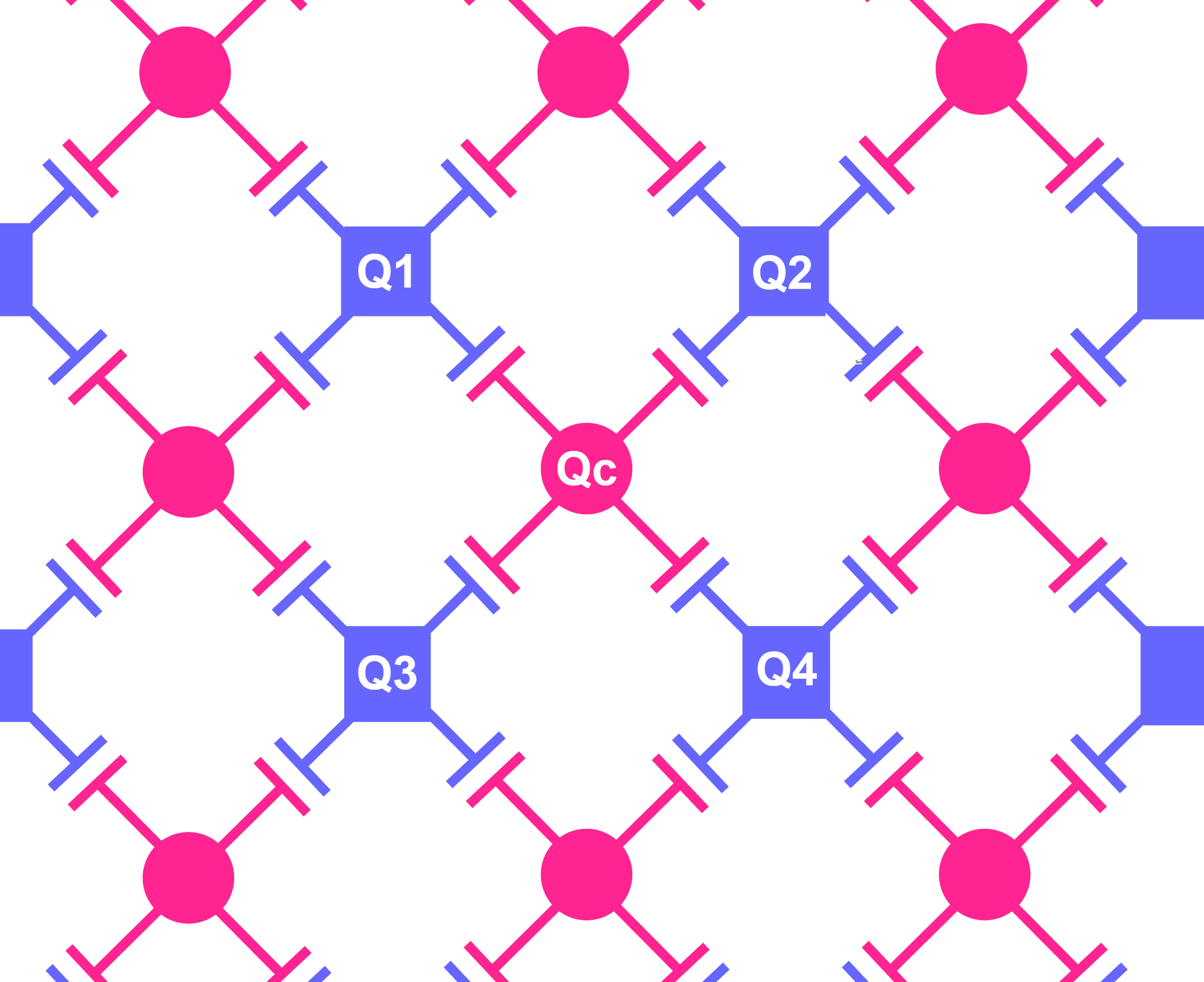}
    \caption{The proposed hybrid coupling topology for maximizing processor scalability. By utilizing the $Q_1$--$Q_c$--$Q_2$ architecture from Fig. 1 as a fundamental unit cell, the system is expanded into a 2D lattice. The blue squares denote fixed-frequency computational transmon qubits ($Q_1$ to $Q_4$), while the pink circles denote flux-tunable transmon couplers ($Q_c$). The interleaved geometry provides high connectivity, where each computational qubit is capacitively coupled to $Q_c$. }
    \label{fig:Circuit level representation of proposed architechture}
\end{figure}

\begin{figure}
    \centering
    \includegraphics[width=1\linewidth]{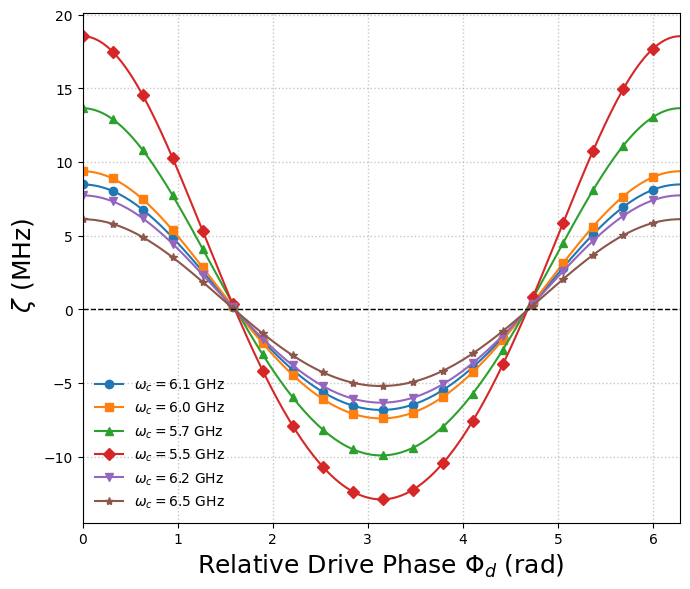}
    \caption{$ZZ$ vs relative drive phase($\Phi_d$) via $Q_c$ (Fig [1]). $\zeta$ is characterized as a function of $\Phi_d$ between off-resonant Stark drives (Fixed amplitude) applied to $Q_1$ and $Q_2$. $Q_c$'s frequency is tuned via external magnetic flux.   }
    \label{fig:zz_vs_phase}
\end{figure}

Driving the control qubit at the target qubit's $|0\rangle$ to $|1\rangle$ transition frequency with $\varepsilon_c$ amplitude, the target qubit experiences an effective drive $\tilde{\varepsilon}_n$  \cite{PhysRevA.100.012301} (depending on the control qubit state $|n\rangle$). This realizes in an entangling $ZX$ interaction with rate $\mu = (\tilde{\varepsilon}_0 \tilde{\varepsilon}_1)/2$.
In the off-resonant regime $\varepsilon_c / \Delta_t \ll 1$, where $\Delta_t = \omega_t - \omega_d$, the target qubit acquires a conditional AC Stark shift
\begin{equation}
\delta_n = \frac{\tilde{\varepsilon}_n^2}{\Delta_t}.
\end{equation}
The resulting $ZZ$ interaction strength, defined as the difference between the Stark shifts, is therefore given by \cite{PhysRevLett.127.200502}
\begin{equation}
\zeta = \delta_0 - \delta_1 = \frac{2\mu(\tilde{\varepsilon}_0 + \tilde{\varepsilon}_1)}{\Delta_t}.
\end{equation}
This is enhanced by applying an additional unconditional drive to the target qubit with amplitude $\varepsilon_t$. To leading order in $\varepsilon_t$, this modifies the effective $ZZ$ interaction as \cite{PhysRevLett.127.200502}
\begin{equation}
\zeta = \frac{2\mu}{\Delta_t}\left(\tilde{\varepsilon}_0 + \tilde{\varepsilon}_1 + 2\varepsilon_t\right) + O(|\varepsilon_t|^2),
\end{equation}
We enhance this topology (fig 1) by adding a tunable transmon between $Q_1$ and $Q_2$.Transmon $Q_c$ is biased so that the detuning $\Delta_{ic}$ is small, which adds a significant boost in Stark-induced interaction shown in Figure  \ref{fig:zz_vs_phase}. This strong interaction can provide a much faster gate than simply deploying a fixed-frequency microwave-tuned two-qubit gate. The interaction Hamiltonian, $
\hat{H}_{\mathrm{int}} = \sum_{k \in \{1,2\}} J_{kc}
\left( \hat{a}_k^\dagger \hat{a}_c + \hat{a}_k \hat{a}_c^\dagger \right)
\
$ is responsible for the nearest-neighbor exchange coupling between the qubits and the coupler. The system is driven by a continuous off-resonant EM field applied to the qubits. One of the key challenges is choosing the qubit parameters before fabrication. There are EM simulation platforms \cite{Choi2023QuantumProAI} for the extraction of the qubit parameters as a function of qubit geometry. 



The numerical simulations were performed using the Quantum Toolbox in Python (QuTip)\cite{LAMBERT20261}. For Leakage effects and higher-order renormalizations, the Hilbert space for each transmon mode is truncated to $N=3$ levels. By numerically diagonalizing the full Hamiltonian $\hat{H}_{\text{total}}$, the static ZZ interaction rate $\zeta_{zz}$ is extracted. 
Maximizing  the wave function overlap ( $|\langle \psi_{\text{eigen}} | ijk_{\text{bare}} \rangle|^2$  ) with the bare tensor product states ( $|i\rangle_{Q1} \otimes |j\rangle_{Q2} \otimes |k\rangle_{C}$ ) , we have indentified the computational basis states ( $\widetilde{|ijk\rangle}$  ).
Taking the assumption that the coupler $Q_c$ is in ground state, the ZZ rate was calculated from the eigenenergies $E_{ijk}$ of the computational subspace ($\zeta_{ZZ} = (E_{110} - E_{100}) - (E_{010} - E_{000})$). The suppression of residual ZZ crosstalk is simulated using QuTip by sweeping the relative phase of the drive tones and the flux-tunable frequency of the coupler across the range $5.5~\text{GHz}$ to $6.5~\text{GHz}$. 

While microwave-driven ZZ tuning \cite{PhysRevLett.127.200502} using only fixed-frequency transmons provides an efficient path to implement two-qubit gates, the design is constrained by microwave crosstalk and frequency crowding as we add more qubits. 
In our hybrid architecture, using a tunable transmon as a mediator coupler provides an additional layer of isolation. Fig \ref{fig:zz_vs_phase} demonstrates a peak tunable $ZZ$ coupling of $\zeta_{max}/2\pi \approx$ 18 MHz, representing a 4x increase over the values reported by just tuning the transition frequency of $Q_c$.  \par

This design approach is maximized by connecting four fixed frequency transmon($Q_1$ to $Q_4$) to the flux tunable central transmon $Q_c$ as shown in Fig. \ref{fig:Circuit level representation of proposed architechture}. For the analysis of the proposed coupling topology with maximized connectivity the analysis are done with parameters $\omega_{1}/2\pi =$ 5.02 GHz, $\omega_{2}/2\pi =$5.23 GHz, $\omega_{3}/2\pi = $5.39 GHz $\omega_{4}/2\pi =$ 5.58 GHz,$\omega_{c(flux)}/2\pi =$ 5.1-6.8 GHz.As the transmons are anharmonic oscillators we have also included the second excited state during analysis, where $\eta_{1}/2\pi =$-207 MHz, $\eta_{2}/2\pi =$-214 MHz, $\eta_{3}/2\pi =$-201 MHz, and $\eta_{4}/2\pi =$-211 MHz. In the drive frame (taking Duffing oscillator approximation), the total Hamiltonian is 

\begin{equation}
\begin{aligned}
H =
&\sum_{i=1}^{4}
\left[
(\omega_i-\omega_d)\, a_i^\dagger a_i
+
\frac{\eta_i}{2}\,
a_i^\dagger a_i^\dagger a_i a_i
\right]
\\[6pt]
&+
\left[
\omega_c(\Phi)\, a_c^\dagger a_c
+
\frac{\eta_c}{2}\,
a_c^\dagger a_c^\dagger a_c a_c
\right]
\\[6pt]
&+
\sum_{i=1}^{4}
J_{ic}
\left(
a_i^\dagger a_c + a_i a_c^\dagger
\right)
\\[6pt]
&+
\sum_{i=1}^{4}
\left(
\varepsilon_i a_i + \varepsilon_i^* a_i^\dagger
\right)
\end{aligned}
\end{equation}

Where $\hat{n}=\hat{a}^{\dagger} \hat{a}$ is the number operator. We have swept the relative drive phase to visualize the effect of drive on $\zeta$ and also analyzed the effect at different drive amplitude ($\epsilon$) and drive frequencies between different pairs.

\begin{figure}
    \centering
    \includegraphics[width=0.9\linewidth]{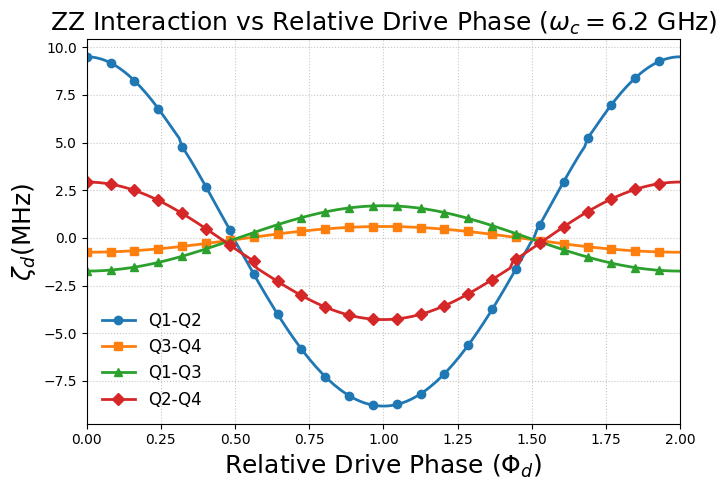}
    \caption{Multi-qubit $ZZ$ interaction for the proposed hybrid coupling topology. $\zeta_d$ is plotted  against $\Phi_d$ for four distinct qubit pairs ($Q_1-Q_2$, $Q_3-Q_4$, $Q_1-Q_3$, $Q_2-Q_4$ when $Q_c$'s first transition ($|0> to |1>$) frequency is tuned to 6.2 GHz}
    \label{fig:placeholder}
\end{figure}

\begin{figure}
    \centering
    \includegraphics[width=1\linewidth]{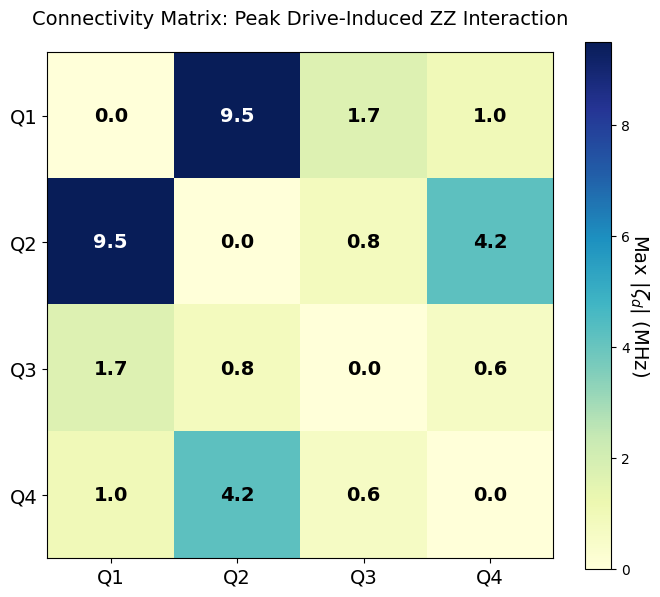}
    \caption{Connectivity matrix analysis of the interaction landscape within the hybrid coupling topology (shown in Fig 6) to capture peak $|\zeta_d|$ for any given pair at fixed off-resonant Stark drive amplitude (when $Q_c$ is tuned to 6.2 GHz  }
    \label{fig:matrix}
\end{figure}

\begin{figure}
    \centering
    \includegraphics[width=1\linewidth]{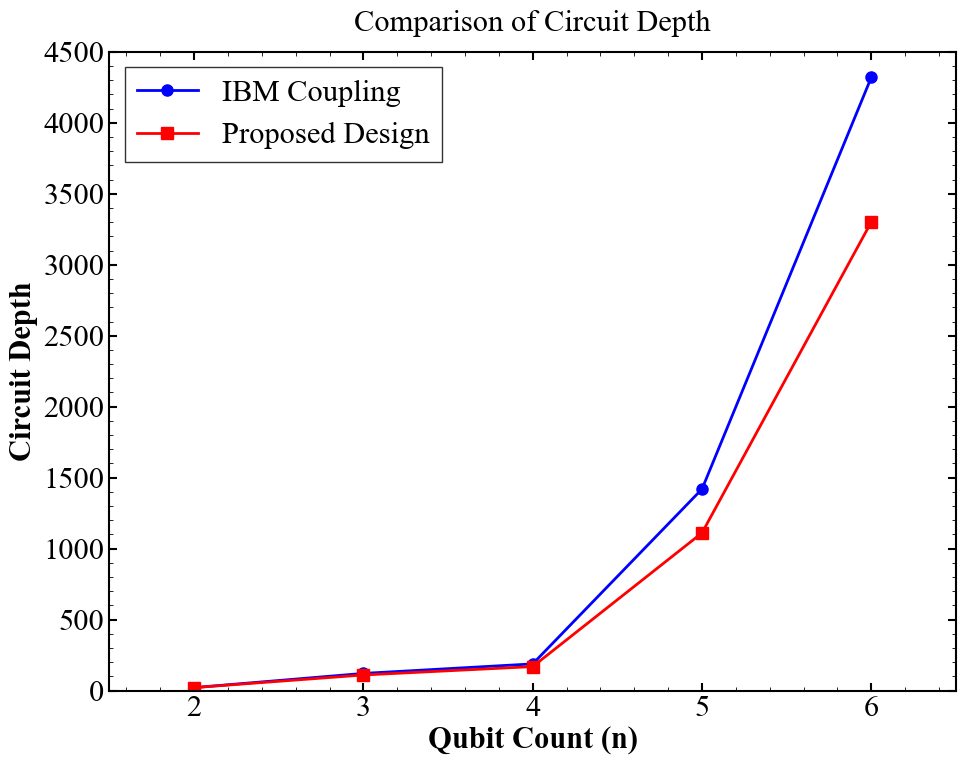}
    \caption{Comparative analysis of circuit depth versus qubit count for Grover's algorithm. The blue line represents the standard IBM coupling, while the red line represents the proposed design. The proposed architecture demonstrates approximately 20\%   lower circuit depth for higher qubit counts +4, indicating improved scalability.}
    \label{fig:depth}
\end{figure}

\begin{figure*}
    \centering
    \includegraphics[width=0.9\linewidth]{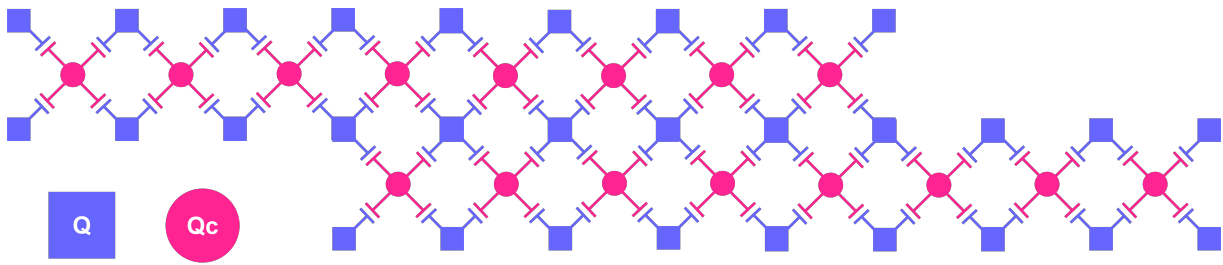}
    \caption{Proposed 2D  grid implementation of the hybrid coupling topology. The grid is composed of fixed-frequency  transmon qubits (squares, $Q$) and flux-tunable transmon couplers ( circles, $Q_c$). }
    \label{fig:full-scale-layout}
\end{figure*}

\begin{figure}
    \centering
    \includegraphics[width=1\linewidth]{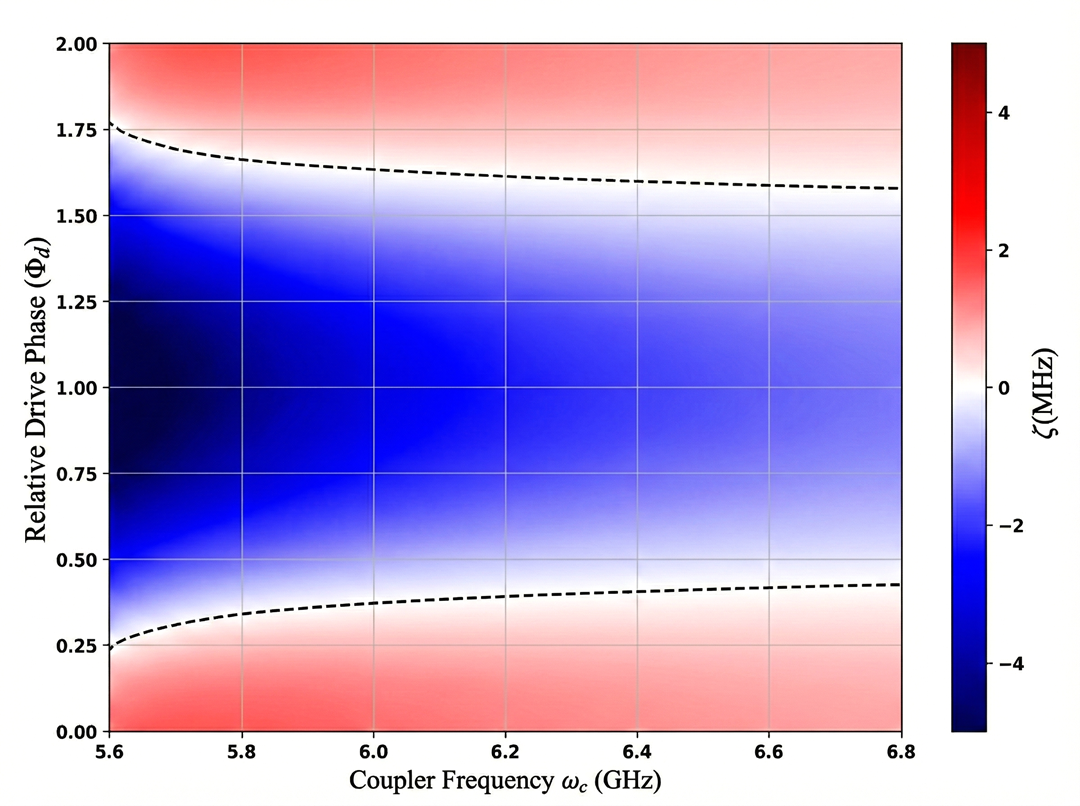}
    \caption{ZZ vs $\omega_c$ and Phase, for the interaction between $Q_1-Q_2$ pair.The sweet spots where ZZ is nearly 0 MHz shown using the dashed black line}
    \label{fig:Proposed layout for Qc Block}
\end{figure}

As the design scales, by repeating the larger unit cell (Fig \ref{fig:Circuit level representation of proposed architechture}), each computational qubit ($Q_n$) sits at the center, can be selectively coupled to its 8 nearest computational qubits.
\section{Results}
To mitigate the always-on static $ZZ$ interactions \cite{PhysRevLett.127.200502} experimentally demonstrated that by applying off-resonant Stark drive at a specific frequency, $\zeta$ can be calibrated by only using fixed frequency transmons.  We have implemented a dual-control scheme combining flux-tunable transmon-mediated interaction with off-resonant Stark drive. As the coupler frequency is tuned from 6.5 GHz down to 5.5GHz (shown in Fig[5]), there is a significant enhancement in interaction strength with a maximum swing of 30 MHz at 5.5 GHz. From fig [5], we can conclude that $\zeta$ is very minimally sensitive to flux fluctuations at zero-crossing, effectively turning off the interaction. The interaction strength can be fine-tuned by adjusting $\omega_c$, where we can compensate for the static $ZZ$. \par
The proposed hybrid topology consisting of four fixed-frequency transmons mediated by a tunable transmon ($Q_c$) overcomes the connectivity constraints of the nearest-neighbor lattices by enabling interactions (mediated by $Q_c$ between any pair of qubits in the cluster. As shown in  Fig[6], the central coupler mediates the interactions across the entire cluster, but the magnitude and relative drive phase sensitivity of $zeta$ are different (pair-specific)  the $\zeta$ for the pair $Q_1 - Q_2$, exhibits a peak interaction strength of $\approx9.5 MHz$ where  $Q_3 -Q_4$ shows a suppressed $\zeta$ at same bias point. The integration of a tunable transmon in this hybrid topology provides a continuously tunable coupling landscape. \par


To verify the reduction in circuit depth of our hybrid coupler, we have utilized IBM's Qiskit SDK. Using Qiskit, it is possible to benchmark the estimated circuit depth against standard coupling topologies under ideal conditions.

\par
 To replicate our proposed design, we have modified the IBM Qiskit FakeBackend and QuTip by overwriting the backend class with our coupling topology. Using this approach, we can simulate the quantum chip locally without fabricating the design.
\par

As a stress test, we have selected Grover's search algorithm for problem sizes ranging from 2 to 6 qubits. This typically incurs high SWAP overhead as we increase the number of qubits. 
 Using the Qiskit transpiler, we have estimated the circuit depth.

Fig [8] shows the circuit depth as a function of the number of qubits for IBM Heron QPU's coupling topology(blue) and our proposed design(orange). It was expected that as the qubit count increases, circuit depth increases exponentially( number of SWAPs increases). Our design outperforms IBM's coupling topology as the qubit count goes beyond four. The proposed design achieves an overall reduction in circuit depth of approximately 20\% over IBM's Heron processor. 
\par

\section{Conclusion}
In this work, we demonstrated an experimentally backed
hybrid tunable coupling cluster which optimizes the trade-
off between scalability and connectivity in a superconducting
quantum processor. By only deploying a single coupler in a
four-qubit cluster, we have demonstrated a hybrid scheme of
coupling any qubit with its nearest four qubits to create
an optimized dense grid with $ZZ$ tunability. The cluster-based
unit cell provides a solution for a large-scale qubit grid
with minimized physical footprint and control hardware. We
have also demonstrated the impact of our proposed design on
circuit depth.

\appendix

\bibliography{apssamp}

\providecommand{\noopsort}[1]{}\providecommand{\singleletter}[1]{#1}%
\begin{thebibliography}{45}%
\makeatletter
\providecommand \@ifxundefined [1]{%
 \@ifx{#1\undefined}
}%
\providecommand \@ifnum [1]{%
 \ifnum #1\expandafter \@firstoftwo
 \else \expandafter \@secondoftwo
 \fi
}%
\providecommand \@ifx [1]{%
 \ifx #1\expandafter \@firstoftwo
 \else \expandafter \@secondoftwo
 \fi
}%
\providecommand \natexlab [1]{#1}%
\providecommand \enquote  [1]{``#1''}%
\providecommand \bibnamefont  [1]{#1}%
\providecommand \bibfnamefont [1]{#1}%
\providecommand \citenamefont [1]{#1}%
\providecommand \href@noop [0]{\@secondoftwo}%
\providecommand \href [0]{\begingroup \@sanitize@url \@href}%
\providecommand \@href[1]{\@@startlink{#1}\@@href}%
\providecommand \@@href[1]{\endgroup#1\@@endlink}%
\providecommand \@sanitize@url [0]{\catcode `\\12\catcode `\$12\catcode
  `\&12\catcode `\#12\catcode `\^12\catcode `\_12\catcode `\%12\relax}%
\providecommand \@@startlink[1]{}%
\providecommand \@@endlink[0]{}%
\providecommand \url  [0]{\begingroup\@sanitize@url \@url }%
\providecommand \@url [1]{\endgroup\@href {#1}{\urlprefix }}%
\providecommand \urlprefix  [0]{URL }%
\providecommand \Eprint [0]{\href }%
\providecommand \doibase [0]{https://doi.org/}%
\providecommand \selectlanguage [0]{\@gobble}%
\providecommand \bibinfo  [0]{\@secondoftwo}%
\providecommand \bibfield  [0]{\@secondoftwo}%
\providecommand \translation [1]{[#1]}%
\providecommand \BibitemOpen [0]{}%
\providecommand \bibitemStop [0]{}%
\providecommand \bibitemNoStop [0]{.\EOS\space}%
\providecommand \EOS [0]{\spacefactor3000\relax}%
\providecommand \BibitemShut  [1]{\csname bibitem#1\endcsname}%
\let\auto@bib@innerbib\@empty
\bibitem [{\citenamefont {Feynman}(1954)}]{feyn54}%
  \BibitemOpen
  \bibfield  {author} {\bibinfo {author} {\bibfnamefont {R.~P.}\ \bibnamefont
  {Feynman}},\ }\href {https://doi.org/10.1029/2002JD002268} {\bibfield
  {journal} {\bibinfo  {journal} {Phys.\ Rev.}\ }\textbf {\bibinfo {volume}
  {94}},\ \bibinfo {pages} {262} (\bibinfo {year} {1954})}\BibitemShut
  {NoStop}%
\bibitem [{\citenamefont {Witten}(2001)}]{witten2001}%
  \BibitemOpen
  \bibfield  {author} {\bibinfo {author} {\bibfnamefont {E.}~\bibnamefont
  {Witten}},\ }\href@noop {} {} (\bibinfo {year} {2001}),\ \Eprint
  {https://arxiv.org/abs/hep-th/0106109} {hep-th/0106109} \BibitemShut
  {NoStop}%
\bibitem [{\citenamefont {Einstein}, \citenamefont {Podolsky},\ and\
  \citenamefont {Rosen}(1935)}]{epr}%
  \BibitemOpen
  \bibfield  {author} {\bibinfo {author} {\bibfnamefont {A.}~\bibnamefont
  {Einstein}}, \bibinfo {author} {\bibfnamefont {{\relax Yu}.}~\bibnamefont
  {Podolsky}},\ and\ \bibinfo {author} {\bibfnamefont {N.}~\bibnamefont
  {Rosen}},\ }\href@noop {} {\bibfield  {journal} {\bibinfo  {journal} {Phys.\
  Rev.}\ }\textbf {\bibinfo {volume} {47}},\ \bibinfo {pages} {777} (\bibinfo
  {year} {1935})}\BibitemShut {NoStop}%
\bibitem [{\citenamefont {Berman}\ and\ \citenamefont
  {Izrailev}(1983)}]{Berman1983}%
  \BibitemOpen
  \bibfield  {author} {\bibinfo {author} {\bibfnamefont {G.~P.}\ \bibnamefont
  {Berman}, \bibfnamefont {Jr.}}\ and\ \bibinfo {author} {\bibfnamefont
  {F.~M.}\ \bibnamefont {Izrailev}, \bibfnamefont {Jr.}},\ }\bibfield  {title}
  {\enquote {\bibinfo {title} {Stability of nonlinear modes},}\ }\href@noop {}
  {\bibfield  {journal} {\bibinfo  {journal} {Physica D}\ }\textbf {\bibinfo
  {volume} {88}},\ \bibinfo {pages} {445} (\bibinfo {year} {1983})}\BibitemShut
  {NoStop}%
\bibitem [{Note1()}]{Note1}%
  \BibitemOpen
  \bibinfo {note} {Automatically placing footnotes into the bibliography
  requires using BibTeX to compile the bibliography.}\BibitemShut {Stop}%
\bibitem [{\citenamefont {Birell}\ and\ \citenamefont {Davies}(1982)}]{ref1}%
  \BibitemOpen
  \bibfield  {author} {\bibinfo {author} {\bibfnamefont {N.~D.}\ \bibnamefont
  {Birell}}\ and\ \bibinfo {author} {\bibfnamefont {P.~C.~W.}\ \bibnamefont
  {Davies}},\ }\href@noop {} {\emph {\bibinfo {title} {Quantum Fields in Curved
  Space}}}\ (\bibinfo  {publisher} {Cambridge University Press},\ \bibinfo
  {year} {1982})\BibitemShut {NoStop}%
\bibitem [{\citenamefont {Davies}\ and\ \citenamefont
  {Parns}(1988)}]{Davies1998}%
  \BibitemOpen
  \bibfield  {author} {\bibinfo {author} {\bibfnamefont {E.~B.}\ \bibnamefont
  {Davies}}\ and\ \bibinfo {author} {\bibfnamefont {L.}~\bibnamefont {Parns}},\
  }\bibfield  {title} {\enquote {\bibinfo {title} {Trapped modes in acoustic
  waveguides},}\ }\href@noop {} {\bibfield  {journal} {\bibinfo  {journal} {Q.
  J. Mech. Appl. Math.}\ }\textbf {\bibinfo {volume} {51}},\ \bibinfo {pages}
  {477--492} (\bibinfo {year} {1988})}\BibitemShut {NoStop}%
\bibitem [{\citenamefont {Beutler}(1994{\natexlab{a}})}]{Beutler1994}%
  \BibitemOpen
  \bibfield  {author} {\bibinfo {author} {\bibfnamefont {E.}~\bibnamefont
  {Beutler}},\ }\enquote {\bibinfo {title} {Williams hematology},}\ \ (\bibinfo
   {publisher} {McGraw-Hill},\ \bibinfo {address} {New York},\ \bibinfo {year}
  {1994})\ Chap.~\bibinfo {chapter} {7}, pp.\ \bibinfo {pages} {654--662},\
  \bibinfo {edition} {5th}\ ed.\BibitemShut {Stop}%
\bibitem [{\citenamefont {Knuth}(1973)}]{inbook-full}%
  \BibitemOpen
  \bibfield  {author} {\bibinfo {author} {\bibfnamefont {D.~E.}\ \bibnamefont
  {Knuth}},\ }\enquote {\bibinfo {title} {Fundamental algorithms},}\ \
  (\bibinfo  {publisher} {Addison-Wesley},\ \bibinfo {address} {Reading,
  Massachusetts},\ \bibinfo {year} {\noopsort{1973b}1973})\ \bibinfo {type}
  {Section}\ \bibinfo {chapter} {1.2}, pp.\ \bibinfo {pages} {10--119},\
  \bibinfo {edition} {2nd}\ ed.,\ \bibinfo {note} {a full INBOOK
  entry}\BibitemShut {NoStop}%
\bibitem [{\citenamefont {Smith}\ and\ \citenamefont
  {Johnson}(2005)}]{Smith2005}%
  \BibitemOpen
  \bibfield  {author} {\bibinfo {author} {\bibfnamefont {J.~S.}\ \bibnamefont
  {Smith}}\ and\ \bibinfo {author} {\bibfnamefont {G.~W.}\ \bibnamefont
  {Johnson}},\ }\href@noop {} {\bibfield  {journal} {\bibinfo  {journal}
  {Philos. Trans. R. Soc. London, Ser. B}\ }\textbf {\bibinfo {volume} {777}},\
  \bibinfo {pages} {1395} (\bibinfo {year} {2005})}\BibitemShut {NoStop}%
\bibitem [{\citenamefont {Smith}, \citenamefont {Johnson},\ and\ \citenamefont
  {Miller}()}]{Smith2010}%
  \BibitemOpen
  \bibfield  {author} {\bibinfo {author} {\bibfnamefont {W.~J.}\ \bibnamefont
  {Smith}}, \bibinfo {author} {\bibfnamefont {T.~J.}\ \bibnamefont {Johnson}},\
  and\ \bibinfo {author} {\bibfnamefont {B.~G.}\ \bibnamefont {Miller}},\
  }\href@noop {} {\enquote {\bibinfo {title} {Surface chemistry and
  preferential crystal orientation on a silicon surface},}\ }\bibinfo {note}
  {{J. Appl. Phys.} (unpublished)}\BibitemShut {NoStop}%
\bibitem [{\citenamefont {Smith}, \citenamefont {Johnson},\ and\ \citenamefont
  {Klein}()}]{Smith2010a}%
  \BibitemOpen
  \bibfield  {author} {\bibinfo {author} {\bibfnamefont {V.~K.}\ \bibnamefont
  {Smith}}, \bibinfo {author} {\bibfnamefont {K.}~\bibnamefont {Johnson}},\
  and\ \bibinfo {author} {\bibfnamefont {M.~O.}\ \bibnamefont {Klein}},\
  }\href@noop {} {\enquote {\bibinfo {title} {Surface chemistry and
  preferential crystal orientation on a silicon surface},}\ }\bibinfo {note}
  {{J. Appl. Phys.} (submitted)}\BibitemShut {NoStop}%
\bibitem [{\citenamefont {{\"{U}}nderwood}, \citenamefont {{\~N}et},\ and\
  \citenamefont {{\={P}}ot}(1988)}]{unpublished-full}%
  \BibitemOpen
  \bibfield  {author} {\bibinfo {author} {\bibfnamefont {U.}~\bibnamefont
  {{\"{U}}nderwood}}, \bibinfo {author} {\bibfnamefont {N.}~\bibnamefont
  {{\~N}et}},\ and\ \bibinfo {author} {\bibfnamefont {P.}~\bibnamefont
  {{\={P}}ot}},\ }\href@noop {} {\enquote {\bibinfo {title} {Lower bounds for
  wishful research results},}\ } (\bibinfo {year} {1988}),\ \bibinfo {note}
  {talk at Fanstord University (A full UNPUBLISHED entry)}\BibitemShut
  {NoStop}%
\bibitem [{\citenamefont {Johnson}, \citenamefont {Miller},\ and\ \citenamefont
  {Smith}(2007)}]{JohnsonMillerSmith2007}%
  \BibitemOpen
  \bibfield  {author} {\bibinfo {author} {\bibfnamefont {M.~P.}\ \bibnamefont
  {Johnson}}, \bibinfo {author} {\bibfnamefont {K.~L.}\ \bibnamefont
  {Miller}},\ and\ \bibinfo {author} {\bibfnamefont {K.}~\bibnamefont
  {Smith}},\ }\href@noop {} {}\bibinfo {howpublished} {personal communication}
  (\bibinfo {year} {2007})\BibitemShut {NoStop}%
\bibitem [{\citenamefont {Smith}(2007{\natexlab{a}})}]{Smith2007}%
  \BibitemOpen
  \bibinfo {editor} {\bibfnamefont {J.}~\bibnamefont {Smith}},\ ed.,\
  \href@noop {} {\emph {\bibinfo {title} {AIP Conf. Proc.}}},\ Vol.\ \bibinfo
  {volume} {841}\ (\bibinfo {year} {2007})\BibitemShut {NoStop}%
\bibitem [{\citenamefont {Oz}\ and\ \citenamefont
  {Yannakakis}(1983)}]{proceedings-full}%
  \BibitemOpen
  \bibinfo {editor} {\bibfnamefont {W.~V.}\ \bibnamefont {Oz}}\ and\ \bibinfo
  {editor} {\bibfnamefont {M.}~\bibnamefont {Yannakakis}},\ eds.,\ \href@noop
  {} {\emph {\bibinfo {title} {Proc. Fifteenth Annual}}},\ \bibinfo {series}
  {All ACM Conferences}\ No.~\bibinfo {number} {17},\ \bibinfo {organization}
  {ACM}\ (\bibinfo  {publisher} {Academic Press},\ \bibinfo {address}
  {Boston},\ \bibinfo {year} {1983})\ \bibinfo {note} {a full PROCEEDINGS
  entry}\BibitemShut {NoStop}%
\bibitem [{\citenamefont {Burstyn}(2004)}]{Burstyn2004}%
  \BibitemOpen
  \bibfield  {author} {\bibinfo {author} {\bibfnamefont {Y.}~\bibnamefont
  {Burstyn}},\ }\href@noop {} {\enquote {\bibinfo {title} {{Proceedings of the
  5th International Molecular Beam Epitaxy Conference, Santa Fe, NM}},}\ }
  (\bibinfo {year} {2004}),\ \bibinfo {note} {(unpublished)}\BibitemShut
  {NoStop}%
\bibitem [{\citenamefont {Quinn}(2001)}]{Quinn2001}%
  \BibitemOpen
  \bibinfo {editor} {\bibfnamefont {B.}~\bibnamefont {Quinn}},\ ed.,\
  \href@noop {} {\emph {\bibinfo {title} {{Proceedings of the 2003 Particle
  Accelerator Conference, Portland, OR, 12-16 May 2005}}}}\ (\bibinfo
  {publisher} {Wiley},\ \bibinfo {address} {New York},\ \bibinfo {year}
  {2001})\ \bibinfo {note} {albeit the conference was held in 2005, it was the
  2003 conference, and the proceedings were published in 2001; go
  figure}\BibitemShut {NoStop}%
\bibitem [{\citenamefont {Agarwal}(2001)}]{Agarwal2001}%
  \BibitemOpen
  \bibfield  {author} {\bibinfo {author} {\bibfnamefont {A.~G.}\ \bibnamefont
  {Agarwal}},\ }\bibfield  {title} {\enquote {\bibinfo {title} {{Proceedings of
  the Fifth Low Temperature Conference, Madison, WI, 1999}},}\ }\href@noop {}
  {\bibfield  {journal} {\bibinfo  {journal} {Semiconductors}\ }\textbf
  {\bibinfo {volume} {66}},\ \bibinfo {pages} {1238} (\bibinfo {year}
  {2001})}\BibitemShut {NoStop}%
\bibitem [{\citenamefont {Smith}()}]{SmithDA01}%
  \BibitemOpen
  \bibfield  {author} {\bibinfo {author} {\bibfnamefont {R.}~\bibnamefont
  {Smith}},\ }\bibfield  {title} {\enquote {\bibinfo {title} {Hummingbirds are
  our friends},}\ }\href@noop {} {\bibfield  {journal} {\bibinfo  {journal} {J.
  Appl. Phys. (these proceedings)}\ }}\bibinfo {note} {Abstract No.
  DA-01}\BibitemShut {NoStop}%
\bibitem [{\citenamefont {Smith}(2007{\natexlab{b}})}]{Smith2007a}%
  \BibitemOpen
  \bibfield  {author} {\bibinfo {author} {\bibfnamefont {J.}~\bibnamefont
  {Smith}},\ }\href@noop {} {\bibfield  {journal} {\bibinfo  {journal} {Proc.
  SPIE}\ }\textbf {\bibinfo {volume} {124}},\ \bibinfo {pages} {367} (\bibinfo
  {year} {2007}{\natexlab{b}})},\ \bibinfo {note} {required title is
  missing}\BibitemShut {NoStop}%
\bibitem [{\citenamefont {T{\'{e}}rrific}(1988)}]{techreport-full}%
  \BibitemOpen
  \bibfield  {author} {\bibinfo {author} {\bibfnamefont {T.}~\bibnamefont
  {T{\'{e}}rrific}},\ }\href@noop {} {\enquote {\bibinfo {title} {An {$O(n \log
  n / \! \log\log n)$} sorting algorithm},}\ }\bibinfo {type} {Wishful Research
  Result}\ \bibinfo {number} {7}\ (\bibinfo  {institution} {Fanstord
  University},\ \bibinfo {address} {Computer Science Department, Fanstord,
  California},\ \bibinfo {year} {1988})\ \bibinfo {note} {a full TECHREPORT
  entry}\BibitemShut {NoStop}%
\bibitem [{\citenamefont {Nelson}(1999{\natexlab{a}})}]{Nelson1999}%
  \BibitemOpen
  \bibfield  {author} {\bibinfo {author} {\bibfnamefont {J.}~\bibnamefont
  {Nelson}},\ }\href@noop {} {}\bibinfo {type} {{TWI Report}}\ \bibinfo
  {number} {666/1999}\ (\bibinfo {year} {Jan.~1999})\ \bibinfo {note} {required
  institution missing}\BibitemShut {NoStop}%
\bibitem [{\citenamefont {Fields}(2005)}]{Fields2005}%
  \BibitemOpen
  \bibfield  {author} {\bibinfo {author} {\bibfnamefont {W.~K.}\ \bibnamefont
  {Fields}},\ }\href@noop {} {}\bibinfo {type} {{ECE Report No.}}\ \bibinfo
  {number} {AL944}\ (\bibinfo {year} {2005})\ \bibinfo {note} {required
  institution missing}\BibitemShut {NoStop}%
\bibitem [{\citenamefont {Zalkins}(2008)}]{Zalkins2008}%
  \BibitemOpen
  \bibfield  {author} {\bibinfo {author} {\bibfnamefont {Y.~M.}\ \bibnamefont
  {Zalkins}},\ }\href@noop {} {}\bibinfo {howpublished} {e-print
  arXiv:cond-mat/040426} (\bibinfo {year} {2008})\BibitemShut {NoStop}%
\bibitem [{\citenamefont {Nelson}(2005)}]{Nelson2005}%
  \BibitemOpen
  \bibfield  {author} {\bibinfo {author} {\bibfnamefont {J.}~\bibnamefont
  {Nelson}},\ }\href@noop {} {}\bibinfo {howpublished} {{U.S. Patent No.}
  5,693,000} (\bibinfo {year} {12~Dec.~2005})\BibitemShut {NoStop}%
\bibitem [{\citenamefont {Nelson}(1999{\natexlab{b}})}]{Nelson1999a}%
  \BibitemOpen
  \bibfield  {author} {\bibinfo {author} {\bibfnamefont {J.~K.}\ \bibnamefont
  {Nelson}},\ }\href@noop {} {\bibinfo {type} {M.{S}. thesis}},\ \bibinfo
  {school} {New York University} (\bibinfo {year}
  {1999}{\natexlab{b}})\BibitemShut {NoStop}%
\bibitem [{\citenamefont {Masterly}(1988)}]{mastersthesis-full}%
  \BibitemOpen
  \bibfield  {author} {\bibinfo {author} {\bibfnamefont {{\'{E}}.}~\bibnamefont
  {Masterly}},\ }\emph {\bibinfo {title} {Mastering Thesis Writing}},\
  \href@noop {} {\bibinfo {type} {Master's project}},\ \bibinfo  {school}
  {Stanford University}, \bibinfo {address} {English Department} (\bibinfo
  {year} {1988}),\ \bibinfo {note} {a full MASTERSTHESIS entry}\BibitemShut
  {NoStop}%
\bibitem [{\citenamefont {Smith}(2003)}]{Smith2003}%
  \BibitemOpen
  \bibfield  {author} {\bibinfo {author} {\bibfnamefont {S.~M.}\ \bibnamefont
  {Smith}},\ }\href@noop {} {\bibinfo {type} {{Ph.D.} thesis}},\ \bibinfo
  {school} {Massachusetts Institute of Technology} (\bibinfo {year}
  {2003})\BibitemShut {NoStop}%
\bibitem [{\citenamefont {Kawa}\ and\ \citenamefont {Lin}(2003)}]{KawaLin2003}%
  \BibitemOpen
  \bibfield  {author} {\bibinfo {author} {\bibfnamefont {S.~R.}\ \bibnamefont
  {Kawa}}\ and\ \bibinfo {author} {\bibfnamefont {S.-J.}\ \bibnamefont {Lin}},\
  }\href@noop {} {\bibfield  {journal} {\bibinfo  {journal} {J. Geophys. Res.}\
  }\textbf {\bibinfo {volume} {108}},\ \bibinfo {pages} {4201} (\bibinfo {year}
  {2003})},\ \bibinfo {note} {{DOI:10.1029/2002JD002268}}\BibitemShut {NoStop}%
\bibitem [{\citenamefont {Phony-Baloney}(1988)}]{phdthesis-full}%
  \BibitemOpen
  \bibfield  {author} {\bibinfo {author} {\bibfnamefont {F.~P.}\ \bibnamefont
  {Phony-Baloney}},\ }\emph {\bibinfo {title} {Fighting Fire with Fire:
  Festooning {F}rench Phrases}},\ \href@noop {} {\bibinfo {type} {{PhD}
  dissertation}},\ \bibinfo  {school} {Fanstord University}, \bibinfo {address}
  {Department of French} (\bibinfo {year} {1988}),\ \bibinfo {note} {a full
  PHDTHESIS entry}\BibitemShut {NoStop}%
\bibitem [{\citenamefont {Knuth}(1981)}]{book-full}%
  \BibitemOpen
  \bibfield  {author} {\bibinfo {author} {\bibfnamefont {D.~E.}\ \bibnamefont
  {Knuth}},\ }\href@noop {} {\emph {\bibinfo {title} {Seminumerical
  Algorithms}}},\ \bibinfo {edition} {2nd}\ ed.,\ \bibinfo {series} {The Art of
  Computer Programming}, Vol.~\bibinfo {volume} {2}\ (\bibinfo  {publisher}
  {Addison-Wesley},\ \bibinfo {address} {Reading, Massachusetts},\ \bibinfo
  {year} {\noopsort{1973c}1981})\ \bibinfo {note} {a full BOOK
  entry}\BibitemShut {NoStop}%
\bibitem [{\citenamefont {Knvth}(1988)}]{booklet-full}%
  \BibitemOpen
  \bibfield  {author} {\bibinfo {author} {\bibfnamefont {J.~C.}\ \bibnamefont
  {Knvth}},\ }\href@noop {} {\enquote {\bibinfo {title} {The programming of
  computer art},}\ }\bibinfo {howpublished} {Vernier Art Center},\ \bibinfo
  {address} {Stanford, California} (\bibinfo {year} {1988}),\ \bibinfo {note}
  {a full BOOKLET entry}\BibitemShut {NoStop}%
\bibitem [{\citenamefont {Ballagh}\ and\ \citenamefont
  {Savage}(2000{\natexlab{a}})}]{ballagh2000}%
  \BibitemOpen
  \bibfield  {author} {\bibinfo {author} {\bibfnamefont {R.}~\bibnamefont
  {Ballagh}}\ and\ \bibinfo {author} {\bibfnamefont {C.}~\bibnamefont
  {Savage}},\ }\enquote {\bibinfo {title} {Bose-einstein condensation: from
  atomic physics to quantum fluids, proceedings of the 13th physics summer
  school},}\ \ (\bibinfo  {publisher} {World Scientific},\ \bibinfo {address}
  {Singapore},\ \bibinfo {year} {2000})\ \Eprint
  {https://arxiv.org/abs/cond-mat/0008070} {cond-mat/0008070} \BibitemShut
  {NoStop}%
\bibitem [{\citenamefont {Opechowski}\ and\ \citenamefont
  {Guccione}()}]{Magnetism}%
  \BibitemOpen
  \bibfield  {author} {\bibinfo {author} {\bibfnamefont {W.}~\bibnamefont
  {Opechowski}}\ and\ \bibinfo {author} {\bibfnamefont {R.}~\bibnamefont
  {Guccione}},\ }\enquote {\bibinfo {title} {Introduction to the theory of
  normal metals},}\ in\ \href@noop {} {\emph {\bibinfo {booktitle}
  {Magnetism}}},\ Vol.\ \bibinfo {volume} {IIa},\ \bibinfo {editor} {edited by\
  \bibinfo {editor} {\bibfnamefont {G.~T.}\ \bibnamefont {Rado}}\ and\ \bibinfo
  {editor} {\bibfnamefont {H.}~\bibnamefont {Suhl}}}\ (\bibinfo  {publisher}
  {Academic Press},\ \bibinfo {address} {New York})\ p.\ \bibinfo {pages}
  {105}\BibitemShut {NoStop}%
\bibitem [{\citenamefont {Opechowski}\ and\ \citenamefont
  {Guccione}(1965{\natexlab{a}})}]{Magnetismb}%
  \BibitemOpen
  \bibfield  {author} {\bibinfo {author} {\bibfnamefont {W.}~\bibnamefont
  {Opechowski}}\ and\ \bibinfo {author} {\bibfnamefont {R.}~\bibnamefont
  {Guccione}},\ }\bibfield  {title} {\enquote {\bibinfo {title} {Introduction
  to the theory of normal metals},}\ }in\ \href@noop {} {\emph {\bibinfo
  {booktitle} {Magnetism}}},\ Vol.\ \bibinfo {volume} {IIa},\ \bibinfo {editor}
  {edited by\ \bibinfo {editor} {\bibfnamefont {G.~T.}\ \bibnamefont {Rado}}\
  and\ \bibinfo {editor} {\bibfnamefont {H.}~\bibnamefont {Suhl}}}\ (\bibinfo
  {publisher} {Academic Press},\ \bibinfo {address} {New York},\ \bibinfo
  {year} {1965})\ p.\ \bibinfo {pages} {105}\BibitemShut {NoStop}%
\bibitem [{\citenamefont {Smith}(1980{\natexlab{a}})}]{Smith80}%
  \BibitemOpen
  \bibfield  {author} {\bibinfo {author} {\bibfnamefont {J.~M.}\ \bibnamefont
  {Smith}},\ }\enquote {\bibinfo {title} {Molecular dynamics},}\ \ (\bibinfo
  {publisher} {Academic},\ \bibinfo {address} {New York},\ \bibinfo {year}
  {1980})\BibitemShut {NoStop}%
\bibitem [{\citenamefont {Zakharov}\ and\ \citenamefont {Shabat}(1971)}]{ZS71}%
  \BibitemOpen
  \bibfield  {author} {\bibinfo {author} {\bibfnamefont {V.~E.}\ \bibnamefont
  {Zakharov}}\ and\ \bibinfo {author} {\bibfnamefont {A.~B.}\ \bibnamefont
  {Shabat}},\ }\bibfield  {title} {\enquote {\bibinfo {title} {Exact theory of
  two-dimensional self-focusing and one-dimensional self-modulation of waves in
  nonlinear media},}\ }\href@noop {} {\bibfield  {journal} {\bibinfo  {journal}
  {Zh. Eksp. Teor. Fiz.}\ }\textbf {\bibinfo {volume} {61}},\ \bibinfo {pages}
  {118--134} (\bibinfo {year} {1971})},\ \translation{Sov. Phys. JETP
  \textbf{34}, 62 (1972)}\BibitemShut {NoStop}%
\bibitem [{\citenamefont {Beutler}(1994{\natexlab{b}})}]{Beutler1994a}%
  \BibitemOpen
  \bibfield  {author} {\bibinfo {author} {\bibfnamefont {E.}~\bibnamefont
  {Beutler}},\ }in\ \href@noop {} {\emph {\bibinfo {booktitle} {Williams
  Hematology}}},\ Vol.~\bibinfo {volume} {2},\ \bibinfo {editor} {edited by\
  \bibinfo {editor} {\bibfnamefont {E.}~\bibnamefont {Beutler}}, \bibinfo
  {editor} {\bibfnamefont {M.~A.}\ \bibnamefont {Lichtman}}, \bibinfo {editor}
  {\bibfnamefont {B.~W.}\ \bibnamefont {Coller}},\ and\ \bibinfo {editor}
  {\bibfnamefont {T.~S.}\ \bibnamefont {Kipps}}}\ (\bibinfo  {publisher}
  {McGraw-Hill},\ \bibinfo {address} {New York},\ \bibinfo {year} {1994})\
  \bibinfo {edition} {5th}\ ed.,\ Chap.~\bibinfo {chapter} {7}, pp.\ \bibinfo
  {pages} {654--662}\BibitemShut {NoStop}%
\bibitem [{\citenamefont {Ballagh}\ and\ \citenamefont
  {Savage}(2000{\natexlab{b}})}]{ballagh2000a}%
  \BibitemOpen
  \bibfield  {author} {\bibinfo {author} {\bibfnamefont {R.}~\bibnamefont
  {Ballagh}}\ and\ \bibinfo {author} {\bibfnamefont {C.}~\bibnamefont
  {Savage}},\ }\bibfield  {title} {\enquote {\bibinfo {title} {Bose-einstein
  condensation: from atomic physics to quantum fluids},}\ }in\ \href@noop {}
  {\emph {\bibinfo {booktitle} {Proceedings of the 13th Physics Summer
  School}}},\ \bibinfo {editor} {edited by\ \bibinfo {editor} {\bibfnamefont
  {C.}~\bibnamefont {Savage}}\ and\ \bibinfo {editor} {\bibfnamefont
  {M.}~\bibnamefont {Das}}}\ (\bibinfo  {publisher} {World Scientific},\
  \bibinfo {address} {Singapore},\ \bibinfo {year} {2000})\ \Eprint
  {https://arxiv.org/abs/cond-mat/0008070} {cond-mat/0008070} \BibitemShut
  {NoStop}%
\bibitem [{\citenamefont {Opechowski}\ and\ \citenamefont
  {Guccione}(1965{\natexlab{b}})}]{Magnetisma}%
  \BibitemOpen
  \bibfield  {author} {\bibinfo {author} {\bibfnamefont {W.}~\bibnamefont
  {Opechowski}}\ and\ \bibinfo {author} {\bibfnamefont {R.}~\bibnamefont
  {Guccione}},\ }\bibfield  {title} {\enquote {\bibinfo {title} {Introduction
  to the theory of normal metals},}\ }in\ \href@noop {} {\emph {\bibinfo
  {booktitle} {Magnetism}}},\ Vol.\ \bibinfo {volume} {IIa},\ \bibinfo {editor}
  {edited by\ \bibinfo {editor} {\bibfnamefont {G.~T.}\ \bibnamefont {Rado}}\
  and\ \bibinfo {editor} {\bibfnamefont {H.}~\bibnamefont {Suhl}}}\ (\bibinfo
  {publisher} {Academic Press},\ \bibinfo {address} {New York},\ \bibinfo
  {year} {1965})\ p.\ \bibinfo {pages} {105}\BibitemShut {NoStop}%
\bibitem [{\citenamefont {Smith}(1980{\natexlab{b}})}]{Smith80a}%
  \BibitemOpen
  \bibfield  {author} {\bibinfo {author} {\bibfnamefont {J.~M.}\ \bibnamefont
  {Smith}},\ }in\ \href@noop {} {\emph {\bibinfo {booktitle} {Molecular
  Dynamics}}},\ \bibinfo {editor} {edited by\ \bibinfo {editor} {\bibfnamefont
  {C.}~\bibnamefont {Brown}}}\ (\bibinfo  {publisher} {Academic},\ \bibinfo
  {address} {New York},\ \bibinfo {year} {1980})\BibitemShut {NoStop}%
\bibitem [{\citenamefont {Lincoll}(1977)}]{incollection-full}%
  \BibitemOpen
  \bibfield  {author} {\bibinfo {author} {\bibfnamefont {D.~D.}\ \bibnamefont
  {Lincoll}},\ }\bibfield  {title} {\enquote {\bibinfo {title} {Semigroups of
  recurrences},}\ }in\ \href@noop {} {\emph {\bibinfo {booktitle} {High Speed
  Computer and Algorithm Organization}}},\ \bibinfo {series and number}
  {\bibinfo {series} {Fast Computers}\ No.~\bibinfo {number} {23}},\ \bibinfo
  {editor} {edited by\ \bibinfo {editor} {\bibfnamefont {D.~J.}\ \bibnamefont
  {Lipcoll}}, \bibinfo {editor} {\bibfnamefont {D.~H.}\ \bibnamefont
  {Lawrie}},\ and\ \bibinfo {editor} {\bibfnamefont {A.~H.}\ \bibnamefont
  {Sameh}}}\ (\bibinfo  {publisher} {Academic Press},\ \bibinfo {address} {New
  York},\ \bibinfo {year} {1977})\ \bibinfo {edition} {3rd}\ ed.,\ \bibinfo
  {type} {Part}~\bibinfo {chapter} {3}, pp.\ \bibinfo {pages} {179--183},\
  \bibinfo {note} {a full INCOLLECTION entry}\BibitemShut {NoStop}%
\bibitem [{\citenamefont {Oaho}, \citenamefont {Ullman},\ and\ \citenamefont
  {Yannakakis}(1983)}]{inproceedings-full}%
  \BibitemOpen
  \bibfield  {author} {\bibinfo {author} {\bibfnamefont {A.~V.}\ \bibnamefont
  {Oaho}}, \bibinfo {author} {\bibfnamefont {J.~D.}\ \bibnamefont {Ullman}},\
  and\ \bibinfo {author} {\bibfnamefont {M.}~\bibnamefont {Yannakakis}},\
  }\bibfield  {title} {\enquote {\bibinfo {title} {On notions of information
  transfer in {VLSI} circuits},}\ }in\ \href@noop {} {\emph {\bibinfo
  {booktitle} {Proc. Fifteenth Annual ACM}}},\ \bibinfo {series and number}
  {\bibinfo {series} {All ACM Conferences}\ No.~\bibinfo {number} {17}},\
  \bibinfo {editor} {edited by\ \bibinfo {editor} {\bibfnamefont {W.~V.}\
  \bibnamefont {Oz}}\ and\ \bibinfo {editor} {\bibfnamefont {M.}~\bibnamefont
  {Yannakakis}}},\ \bibinfo {organization} {ACM}\ (\bibinfo  {publisher}
  {Academic Press},\ \bibinfo {address} {Boston},\ \bibinfo {year} {1983})\
  pp.\ \bibinfo {pages} {133--139},\ \bibinfo {note} {a full INPROCEDINGS
  entry}\BibitemShut {NoStop}%
\bibitem [{\citenamefont {Manmaker}(1986)}]{manual-full}%
  \BibitemOpen
  \bibfield  {author} {\bibinfo {author} {\bibfnamefont {L.}~\bibnamefont
  {Manmaker}},\ }\href@noop {} {\emph {\bibinfo {title} {The Definitive
  Computer Manual}}},\ \bibinfo {organization} {Chips-R-Us},\ \bibinfo
  {address} {Silicon Valley},\ \bibinfo {edition} {silver}\ ed. (\bibinfo
  {year} {1986}),\ \bibinfo {note} {a full MANUAL entry}\BibitemShut {NoStop}%
\end{thebibliography}%


\begin{thebibliography}{16}%
\makeatletter
\providecommand \@ifxundefined [1]{%
 \@ifx{#1\undefined}
}%
\providecommand \@ifnum [1]{%
 \ifnum #1\expandafter \@firstoftwo
 \else \expandafter \@secondoftwo
 \fi
}%
\providecommand \@ifx [1]{%
 \ifx #1\expandafter \@firstoftwo
 \else \expandafter \@secondoftwo
 \fi
}%
\providecommand \natexlab [1]{#1}%
\providecommand \enquote  [1]{``#1''}%
\providecommand \bibnamefont  [1]{#1}%
\providecommand \bibfnamefont [1]{#1}%
\providecommand \citenamefont [1]{#1}%
\providecommand \href@noop [0]{\@secondoftwo}%
\providecommand \href [0]{\begingroup \@sanitize@url \@href}%
\providecommand \@href[1]{\@@startlink{#1}\@@href}%
\providecommand \@@href[1]{\endgroup#1\@@endlink}%
\providecommand \@sanitize@url [0]{\catcode `\\12\catcode `\$12\catcode
  `\&12\catcode `\#12\catcode `\^12\catcode `\_12\catcode `\%12\relax}%
\providecommand \@@startlink[1]{}%
\providecommand \@@endlink[0]{}%
\providecommand \url  [0]{\begingroup\@sanitize@url \@url }%
\providecommand \@url [1]{\endgroup\@href {#1}{\urlprefix }}%
\providecommand \urlprefix  [0]{URL }%
\providecommand \Eprint [0]{\href }%
\providecommand \doibase [0]{https://doi.org/}%
\providecommand \selectlanguage [0]{\@gobble}%
\providecommand \bibinfo  [0]{\@secondoftwo}%
\providecommand \bibfield  [0]{\@secondoftwo}%
\providecommand \translation [1]{[#1]}%
\providecommand \BibitemOpen [0]{}%
\providecommand \bibitemStop [0]{}%
\providecommand \bibitemNoStop [0]{.\EOS\space}%
\providecommand \EOS [0]{\spacefactor3000\relax}%
\providecommand \BibitemShut  [1]{\csname bibitem#1\endcsname}%
\let\auto@bib@innerbib\@empty
\bibitem [{\citenamefont {Nakamura}\ \emph {et~al.}(1999)\citenamefont
  {Nakamura}, \citenamefont {Pashkin},\ and\ \citenamefont
  {Tsai}}]{Nakamura1999CoherentCO}%
  \BibitemOpen
  \bibfield  {author} {\bibinfo {author} {\bibfnamefont {Y.}~\bibnamefont
  {Nakamura}}, \bibinfo {author} {\bibfnamefont {Y.~A.}\ \bibnamefont
  {Pashkin}},\ and\ \bibinfo {author} {\bibfnamefont {J.~S.}\ \bibnamefont
  {Tsai}},\ }\bibfield  {title} {\bibinfo {title} {Coherent control of
  macroscopic quantum states in a single-cooper-pair box},\ }\href
  {https://doi.org/10.1038/19718} {\bibfield  {journal} {\bibinfo  {journal}
  {Nature}\ }\textbf {\bibinfo {volume} {398}},\ \bibinfo {pages} {786}
  (\bibinfo {year} {1999})}\BibitemShut {NoStop}%
\bibitem [{\citenamefont {Tuokkola}\ \emph {et~al.}(2024)\citenamefont
  {Tuokkola}, \citenamefont {Sunada}, \citenamefont {Kivij{\"a}rvi},
  \citenamefont {Albanese}, \citenamefont {Gr{\"o}nberg}, \citenamefont
  {Kaikkonen}, \citenamefont {Vesterinen}, \citenamefont {Govenius},\ and\
  \citenamefont {M{\"o}tt{\"o}nen}}]{REF1}%
  \BibitemOpen
  \bibfield  {author} {\bibinfo {author} {\bibfnamefont {M.}~\bibnamefont
  {Tuokkola}}, \bibinfo {author} {\bibfnamefont {Y.}~\bibnamefont {Sunada}},
  \bibinfo {author} {\bibfnamefont {H.}~\bibnamefont {Kivij{\"a}rvi}}, \bibinfo
  {author} {\bibfnamefont {J.}~\bibnamefont {Albanese}}, \bibinfo {author}
  {\bibfnamefont {L.}~\bibnamefont {Gr{\"o}nberg}}, \bibinfo {author}
  {\bibfnamefont {J.-P.}\ \bibnamefont {Kaikkonen}}, \bibinfo {author}
  {\bibfnamefont {V.}~\bibnamefont {Vesterinen}}, \bibinfo {author}
  {\bibfnamefont {J.}~\bibnamefont {Govenius}},\ and\ \bibinfo {author}
  {\bibfnamefont {M.}~\bibnamefont {M{\"o}tt{\"o}nen}},\ }\bibfield  {title}
  {\bibinfo {title} {Methods to achieve near-millisecond energy relaxation and
  dephasing times for a superconducting transmon qubit},\ }\href
  {https://api.semanticscholar.org/CorpusID:271516179} {\bibfield  {journal}
  {\bibinfo  {journal} {Nature Communications}\ }\textbf {\bibinfo {volume}
  {16}} (\bibinfo {year} {2024})}\BibitemShut {NoStop}%
\bibitem [{\citenamefont {Yan}\ \emph {et~al.}(2018)\citenamefont {Yan},
  \citenamefont {Krantz}, \citenamefont {Sung}, \citenamefont {Kjaergaard},
  \citenamefont {Campbell}, \citenamefont {Orlando}, \citenamefont
  {Gustavsson},\ and\ \citenamefont {Oliver}}]{PhysRevApplied.10.054062}%
  \BibitemOpen
  \bibfield  {author} {\bibinfo {author} {\bibfnamefont {F.}~\bibnamefont
  {Yan}}, \bibinfo {author} {\bibfnamefont {P.}~\bibnamefont {Krantz}},
  \bibinfo {author} {\bibfnamefont {Y.}~\bibnamefont {Sung}}, \bibinfo {author}
  {\bibfnamefont {M.}~\bibnamefont {Kjaergaard}}, \bibinfo {author}
  {\bibfnamefont {D.~L.}\ \bibnamefont {Campbell}}, \bibinfo {author}
  {\bibfnamefont {T.~P.}\ \bibnamefont {Orlando}}, \bibinfo {author}
  {\bibfnamefont {S.}~\bibnamefont {Gustavsson}},\ and\ \bibinfo {author}
  {\bibfnamefont {W.~D.}\ \bibnamefont {Oliver}},\ }\bibfield  {title}
  {\bibinfo {title} {Tunable coupling scheme for implementing high-fidelity
  two-qubit gates},\ }\href {https://doi.org/10.1103/PhysRevApplied.10.054062}
  {\bibfield  {journal} {\bibinfo  {journal} {Phys. Rev. Appl.}\ }\textbf
  {\bibinfo {volume} {10}},\ \bibinfo {pages} {054062} (\bibinfo {year}
  {2018})}\BibitemShut {NoStop}%
\bibitem [{\citenamefont {Geller}\ \emph {et~al.}(2014)\citenamefont {Geller},
  \citenamefont {Donate}, \citenamefont {Chen}, \citenamefont {Neill},
  \citenamefont {Roushan},\ and\ \citenamefont
  {Martinis}}]{Geller2014TunableCF}%
  \BibitemOpen
  \bibfield  {author} {\bibinfo {author} {\bibfnamefont {M.~R.}\ \bibnamefont
  {Geller}}, \bibinfo {author} {\bibfnamefont {E.}~\bibnamefont {Donate}},
  \bibinfo {author} {\bibfnamefont {Y.}~\bibnamefont {Chen}}, \bibinfo {author}
  {\bibfnamefont {C.~J.}\ \bibnamefont {Neill}}, \bibinfo {author}
  {\bibfnamefont {P.}~\bibnamefont {Roushan}},\ and\ \bibinfo {author}
  {\bibfnamefont {J.~M.}\ \bibnamefont {Martinis}},\ }\bibfield  {title}
  {\bibinfo {title} {Tunable coupler for superconducting xmon qubits:
  Perturbative nonlinear model},\ }\href
  {https://api.semanticscholar.org/CorpusID:7616700} {\bibfield  {journal}
  {\bibinfo  {journal} {arXiv: Quantum Physics}\ } (\bibinfo {year}
  {2014})}\BibitemShut {NoStop}%
\bibitem [{\citenamefont {Mitchell}\ \emph {et~al.}(2021)\citenamefont
  {Mitchell}, \citenamefont {Naik}, \citenamefont {Morvan}, \citenamefont
  {Hashim}, \citenamefont {Kreikebaum}, \citenamefont {Marinelli},
  \citenamefont {Lavrijsen}, \citenamefont {Nowrouzi}, \citenamefont
  {Santiago},\ and\ \citenamefont {Siddiqi}}]{PhysRevLett.127.200502}%
  \BibitemOpen
  \bibfield  {author} {\bibinfo {author} {\bibfnamefont {B.~K.}\ \bibnamefont
  {Mitchell}}, \bibinfo {author} {\bibfnamefont {R.~K.}\ \bibnamefont {Naik}},
  \bibinfo {author} {\bibfnamefont {A.}~\bibnamefont {Morvan}}, \bibinfo
  {author} {\bibfnamefont {A.}~\bibnamefont {Hashim}}, \bibinfo {author}
  {\bibfnamefont {J.~M.}\ \bibnamefont {Kreikebaum}}, \bibinfo {author}
  {\bibfnamefont {B.}~\bibnamefont {Marinelli}}, \bibinfo {author}
  {\bibfnamefont {W.}~\bibnamefont {Lavrijsen}}, \bibinfo {author}
  {\bibfnamefont {K.}~\bibnamefont {Nowrouzi}}, \bibinfo {author}
  {\bibfnamefont {D.~I.}\ \bibnamefont {Santiago}},\ and\ \bibinfo {author}
  {\bibfnamefont {I.}~\bibnamefont {Siddiqi}},\ }\bibfield  {title} {\bibinfo
  {title} {Hardware-efficient microwave-activated tunable coupling between
  superconducting qubits},\ }\href
  {https://doi.org/10.1103/PhysRevLett.127.200502} {\bibfield  {journal}
  {\bibinfo  {journal} {Phys. Rev. Lett.}\ }\textbf {\bibinfo {volume} {127}},\
  \bibinfo {pages} {200502} (\bibinfo {year} {2021})}\BibitemShut {NoStop}%
\bibitem [{\citenamefont {Chow}\ \emph {et~al.}(2021)\citenamefont {Chow},
  \citenamefont {Dial},\ and\ \citenamefont {Gambetta}}]{ibm2021eagle}%
  \BibitemOpen
  \bibfield  {author} {\bibinfo {author} {\bibfnamefont {J.}~\bibnamefont
  {Chow}}, \bibinfo {author} {\bibfnamefont {O.}~\bibnamefont {Dial}},\ and\
  \bibinfo {author} {\bibfnamefont {J.}~\bibnamefont {Gambetta}},\ }\href
  {https://www.ibm.com/quantum/blog/127-qubit-quantum-processor-eagle}
  {\bibinfo {title} {Ibm quantum breaks the 100-qubit processor barrier}}
  (\bibinfo {year} {2021}),\ \bibinfo {note} {blog post}\BibitemShut {NoStop}%
\bibitem [{\citenamefont {Arute}\ \emph {et~al.}(2019)\citenamefont {Arute},
  \citenamefont {Arya}, \citenamefont {Babbush}, \citenamefont {Bacon},
  \citenamefont {Bardin}, \citenamefont {Barends}, \citenamefont {Biswas},
  \citenamefont {Boixo}, \citenamefont {Brandao}, \citenamefont {Buell},
  \citenamefont {Burkett}, \citenamefont {Chen}, \citenamefont {Chen},
  \citenamefont {Chiaro}, \citenamefont {Collins}, \citenamefont {Courtney},
  \citenamefont {Dunsworth}, \citenamefont {Farhi}, \citenamefont {Foxen},
  \citenamefont {Fowler}, \citenamefont {Gidney}, \citenamefont {Giustina},
  \citenamefont {Graff}, \citenamefont {Guerin}, \citenamefont {Habegger},
  \citenamefont {Harrigan}, \citenamefont {Hartmann}, \citenamefont {Ho},
  \citenamefont {Hoffmann}, \citenamefont {Huang}, \citenamefont {Humble},
  \citenamefont {Isakov}, \citenamefont {Jeffrey}, \citenamefont {Jiang},
  \citenamefont {Kafri}, \citenamefont {Kechedzhi}, \citenamefont {Kelly},
  \citenamefont {Klimov}, \citenamefont {Knysh}, \citenamefont {Korotkov},
  \citenamefont {Kostritsa}, \citenamefont {Landhuis}, \citenamefont
  {Lindmark}, \citenamefont {Lucero}, \citenamefont {Lyakh}, \citenamefont
  {Mandr{\`a}}, \citenamefont {McClean}, \citenamefont {McEwen}, \citenamefont
  {Megrant}, \citenamefont {Mi}, \citenamefont {Michielsen}, \citenamefont
  {Mohseni}, \citenamefont {Mutus}, \citenamefont {Naaman}, \citenamefont
  {Neeley}, \citenamefont {Neill}, \citenamefont {Niu}, \citenamefont {Ostby},
  \citenamefont {Petukhov}, \citenamefont {Platt}, \citenamefont {Quintana},
  \citenamefont {Rieffel}, \citenamefont {Roushan}, \citenamefont {Rubin},
  \citenamefont {Sank}, \citenamefont {Satzinger}, \citenamefont {Smelyanskiy},
  \citenamefont {Sung}, \citenamefont {Trevithick}, \citenamefont
  {Vainsencher}, \citenamefont {Villalonga}, \citenamefont {White},
  \citenamefont {Yao}, \citenamefont {Yeh}, \citenamefont {Zalcman},
  \citenamefont {Neven},\ and\ \citenamefont {Martinis}}]{Arute2019}%
  \BibitemOpen
  \bibfield  {author} {\bibinfo {author} {\bibfnamefont {F.}~\bibnamefont
  {Arute}}, \bibinfo {author} {\bibfnamefont {K.}~\bibnamefont {Arya}},
  \bibinfo {author} {\bibfnamefont {R.}~\bibnamefont {Babbush}}, \bibinfo
  {author} {\bibfnamefont {D.}~\bibnamefont {Bacon}}, \bibinfo {author}
  {\bibfnamefont {J.~C.}\ \bibnamefont {Bardin}}, \bibinfo {author}
  {\bibfnamefont {R.}~\bibnamefont {Barends}}, \bibinfo {author} {\bibfnamefont
  {R.}~\bibnamefont {Biswas}}, \bibinfo {author} {\bibfnamefont
  {S.}~\bibnamefont {Boixo}}, \bibinfo {author} {\bibfnamefont {F.~G. S.~L.}\
  \bibnamefont {Brandao}}, \bibinfo {author} {\bibfnamefont {D.~A.}\
  \bibnamefont {Buell}}, \bibinfo {author} {\bibfnamefont {B.}~\bibnamefont
  {Burkett}}, \bibinfo {author} {\bibfnamefont {Y.}~\bibnamefont {Chen}},
  \bibinfo {author} {\bibfnamefont {Z.}~\bibnamefont {Chen}}, \bibinfo {author}
  {\bibfnamefont {B.}~\bibnamefont {Chiaro}}, \bibinfo {author} {\bibfnamefont
  {R.}~\bibnamefont {Collins}}, \bibinfo {author} {\bibfnamefont
  {W.}~\bibnamefont {Courtney}}, \bibinfo {author} {\bibfnamefont
  {A.}~\bibnamefont {Dunsworth}}, \bibinfo {author} {\bibfnamefont
  {E.}~\bibnamefont {Farhi}}, \bibinfo {author} {\bibfnamefont
  {B.}~\bibnamefont {Foxen}}, \bibinfo {author} {\bibfnamefont
  {A.}~\bibnamefont {Fowler}}, \bibinfo {author} {\bibfnamefont
  {C.}~\bibnamefont {Gidney}}, \bibinfo {author} {\bibfnamefont
  {M.}~\bibnamefont {Giustina}}, \bibinfo {author} {\bibfnamefont
  {R.}~\bibnamefont {Graff}}, \bibinfo {author} {\bibfnamefont
  {K.}~\bibnamefont {Guerin}}, \bibinfo {author} {\bibfnamefont
  {S.}~\bibnamefont {Habegger}}, \bibinfo {author} {\bibfnamefont {M.~P.}\
  \bibnamefont {Harrigan}}, \bibinfo {author} {\bibfnamefont {M.~J.}\
  \bibnamefont {Hartmann}}, \bibinfo {author} {\bibfnamefont {A.}~\bibnamefont
  {Ho}}, \bibinfo {author} {\bibfnamefont {M.}~\bibnamefont {Hoffmann}},
  \bibinfo {author} {\bibfnamefont {T.}~\bibnamefont {Huang}}, \bibinfo
  {author} {\bibfnamefont {T.~S.}\ \bibnamefont {Humble}}, \bibinfo {author}
  {\bibfnamefont {S.~V.}\ \bibnamefont {Isakov}}, \bibinfo {author}
  {\bibfnamefont {E.}~\bibnamefont {Jeffrey}}, \bibinfo {author} {\bibfnamefont
  {Z.}~\bibnamefont {Jiang}}, \bibinfo {author} {\bibfnamefont
  {D.}~\bibnamefont {Kafri}}, \bibinfo {author} {\bibfnamefont
  {K.}~\bibnamefont {Kechedzhi}}, \bibinfo {author} {\bibfnamefont
  {J.}~\bibnamefont {Kelly}}, \bibinfo {author} {\bibfnamefont {P.~V.}\
  \bibnamefont {Klimov}}, \bibinfo {author} {\bibfnamefont {S.}~\bibnamefont
  {Knysh}}, \bibinfo {author} {\bibfnamefont {A.}~\bibnamefont {Korotkov}},
  \bibinfo {author} {\bibfnamefont {F.}~\bibnamefont {Kostritsa}}, \bibinfo
  {author} {\bibfnamefont {D.}~\bibnamefont {Landhuis}}, \bibinfo {author}
  {\bibfnamefont {M.}~\bibnamefont {Lindmark}}, \bibinfo {author}
  {\bibfnamefont {E.}~\bibnamefont {Lucero}}, \bibinfo {author} {\bibfnamefont
  {D.}~\bibnamefont {Lyakh}}, \bibinfo {author} {\bibfnamefont
  {S.}~\bibnamefont {Mandr{\`a}}}, \bibinfo {author} {\bibfnamefont {J.~R.}\
  \bibnamefont {McClean}}, \bibinfo {author} {\bibfnamefont {M.}~\bibnamefont
  {McEwen}}, \bibinfo {author} {\bibfnamefont {A.}~\bibnamefont {Megrant}},
  \bibinfo {author} {\bibfnamefont {X.}~\bibnamefont {Mi}}, \bibinfo {author}
  {\bibfnamefont {K.}~\bibnamefont {Michielsen}}, \bibinfo {author}
  {\bibfnamefont {M.}~\bibnamefont {Mohseni}}, \bibinfo {author} {\bibfnamefont
  {J.}~\bibnamefont {Mutus}}, \bibinfo {author} {\bibfnamefont
  {O.}~\bibnamefont {Naaman}}, \bibinfo {author} {\bibfnamefont
  {M.}~\bibnamefont {Neeley}}, \bibinfo {author} {\bibfnamefont
  {C.}~\bibnamefont {Neill}}, \bibinfo {author} {\bibfnamefont {M.~Y.}\
  \bibnamefont {Niu}}, \bibinfo {author} {\bibfnamefont {E.}~\bibnamefont
  {Ostby}}, \bibinfo {author} {\bibfnamefont {A.}~\bibnamefont {Petukhov}},
  \bibinfo {author} {\bibfnamefont {J.~C.}\ \bibnamefont {Platt}}, \bibinfo
  {author} {\bibfnamefont {C.}~\bibnamefont {Quintana}}, \bibinfo {author}
  {\bibfnamefont {E.~G.}\ \bibnamefont {Rieffel}}, \bibinfo {author}
  {\bibfnamefont {P.}~\bibnamefont {Roushan}}, \bibinfo {author} {\bibfnamefont
  {N.~C.}\ \bibnamefont {Rubin}}, \bibinfo {author} {\bibfnamefont
  {D.}~\bibnamefont {Sank}}, \bibinfo {author} {\bibfnamefont {K.~J.}\
  \bibnamefont {Satzinger}}, \bibinfo {author} {\bibfnamefont {V.}~\bibnamefont
  {Smelyanskiy}}, \bibinfo {author} {\bibfnamefont {K.~J.}\ \bibnamefont
  {Sung}}, \bibinfo {author} {\bibfnamefont {M.~D.}\ \bibnamefont
  {Trevithick}}, \bibinfo {author} {\bibfnamefont {A.}~\bibnamefont
  {Vainsencher}}, \bibinfo {author} {\bibfnamefont {B.}~\bibnamefont
  {Villalonga}}, \bibinfo {author} {\bibfnamefont {T.}~\bibnamefont {White}},
  \bibinfo {author} {\bibfnamefont {Z.~J.}\ \bibnamefont {Yao}}, \bibinfo
  {author} {\bibfnamefont {P.}~\bibnamefont {Yeh}}, \bibinfo {author}
  {\bibfnamefont {A.}~\bibnamefont {Zalcman}}, \bibinfo {author} {\bibfnamefont
  {H.}~\bibnamefont {Neven}},\ and\ \bibinfo {author} {\bibfnamefont {J.~M.}\
  \bibnamefont {Martinis}},\ }\bibfield  {title} {\bibinfo {title} {Quantum
  supremacy using a programmable superconducting processor},\ }\href
  {https://doi.org/10.1038/s41586-019-1666-5} {\bibfield  {journal} {\bibinfo
  {journal} {Nature}\ }\textbf {\bibinfo {volume} {574}},\ \bibinfo {pages}
  {505} (\bibinfo {year} {2019})}\BibitemShut {NoStop}%
\bibitem [{\citenamefont {Chen}\ \emph {et~al.}(2014)\citenamefont {Chen},
  \citenamefont {Neill}, \citenamefont {Roushan}, \citenamefont {Leung},
  \citenamefont {Fang}, \citenamefont {Barends}, \citenamefont {Kelly},
  \citenamefont {Campbell}, \citenamefont {Chen}, \citenamefont {Chiaro},
  \citenamefont {Dunsworth}, \citenamefont {Jeffrey}, \citenamefont {Megrant},
  \citenamefont {Mutus}, \citenamefont {O'Malley}, \citenamefont {Quintana},
  \citenamefont {Sank}, \citenamefont {Vainsencher}, \citenamefont {Wenner},
  \citenamefont {White}, \citenamefont {Geller}, \citenamefont {Cleland},\ and\
  \citenamefont {Martinis}}]{PhysRevLett.113.220502}%
  \BibitemOpen
  \bibfield  {author} {\bibinfo {author} {\bibfnamefont {Y.}~\bibnamefont
  {Chen}}, \bibinfo {author} {\bibfnamefont {C.}~\bibnamefont {Neill}},
  \bibinfo {author} {\bibfnamefont {P.}~\bibnamefont {Roushan}}, \bibinfo
  {author} {\bibfnamefont {N.}~\bibnamefont {Leung}}, \bibinfo {author}
  {\bibfnamefont {M.}~\bibnamefont {Fang}}, \bibinfo {author} {\bibfnamefont
  {R.}~\bibnamefont {Barends}}, \bibinfo {author} {\bibfnamefont
  {J.}~\bibnamefont {Kelly}}, \bibinfo {author} {\bibfnamefont
  {B.}~\bibnamefont {Campbell}}, \bibinfo {author} {\bibfnamefont
  {Z.}~\bibnamefont {Chen}}, \bibinfo {author} {\bibfnamefont {B.}~\bibnamefont
  {Chiaro}}, \bibinfo {author} {\bibfnamefont {A.}~\bibnamefont {Dunsworth}},
  \bibinfo {author} {\bibfnamefont {E.}~\bibnamefont {Jeffrey}}, \bibinfo
  {author} {\bibfnamefont {A.}~\bibnamefont {Megrant}}, \bibinfo {author}
  {\bibfnamefont {J.~Y.}\ \bibnamefont {Mutus}}, \bibinfo {author}
  {\bibfnamefont {P.~J.~J.}\ \bibnamefont {O'Malley}}, \bibinfo {author}
  {\bibfnamefont {C.~M.}\ \bibnamefont {Quintana}}, \bibinfo {author}
  {\bibfnamefont {D.}~\bibnamefont {Sank}}, \bibinfo {author} {\bibfnamefont
  {A.}~\bibnamefont {Vainsencher}}, \bibinfo {author} {\bibfnamefont
  {J.}~\bibnamefont {Wenner}}, \bibinfo {author} {\bibfnamefont {T.~C.}\
  \bibnamefont {White}}, \bibinfo {author} {\bibfnamefont {M.~R.}\ \bibnamefont
  {Geller}}, \bibinfo {author} {\bibfnamefont {A.~N.}\ \bibnamefont
  {Cleland}},\ and\ \bibinfo {author} {\bibfnamefont {J.~M.}\ \bibnamefont
  {Martinis}},\ }\bibfield  {title} {\bibinfo {title} {Qubit architecture with
  high coherence and fast tunable coupling},\ }\href
  {https://doi.org/10.1103/PhysRevLett.113.220502} {\bibfield  {journal}
  {\bibinfo  {journal} {Phys. Rev. Lett.}\ }\textbf {\bibinfo {volume} {113}},\
  \bibinfo {pages} {220502} (\bibinfo {year} {2014})}\BibitemShut {NoStop}%
\bibitem [{\citenamefont {Li}\ \emph {et~al.}(2023)\citenamefont {Li},
  \citenamefont {Lu}, \citenamefont {Han}, \citenamefont {Qin}, \citenamefont
  {Ju},\ and\ \citenamefont {Liu}}]{Li2023}%
  \BibitemOpen
  \bibfield  {author} {\bibinfo {author} {\bibfnamefont {H.}~\bibnamefont
  {Li}}, \bibinfo {author} {\bibfnamefont {K.}~\bibnamefont {Lu}}, \bibinfo
  {author} {\bibfnamefont {Z.}~\bibnamefont {Han}}, \bibinfo {author}
  {\bibfnamefont {H.}~\bibnamefont {Qin}}, \bibinfo {author} {\bibfnamefont
  {M.}~\bibnamefont {Ju}},\ and\ \bibinfo {author} {\bibfnamefont
  {S.}~\bibnamefont {Liu}},\ }\bibfield  {title} {\bibinfo {title} {Research on
  qubit mapping technique based on batch swap optimization},\ }\bibfield
  {journal} {\bibinfo  {journal} {International Journal of Advanced Computer
  Science and Applications}\ }\textbf {\bibinfo {volume} {14}},\ \href
  {https://doi.org/10.14569/IJACSA.2023.0141217} {10.14569/IJACSA.2023.0141217}
  (\bibinfo {year} {2023})\BibitemShut {NoStop}%
\bibitem [{\citenamefont {Koch}\ \emph {et~al.}(2007)\citenamefont {Koch},
  \citenamefont {Yu}, \citenamefont {Gambetta}, \citenamefont {Houck},
  \citenamefont {Schuster}, \citenamefont {Majer}, \citenamefont {Blais},
  \citenamefont {Devoret}, \citenamefont {Girvin},\ and\ \citenamefont
  {Schoelkopf}}]{PhysRevA.76.042319}%
  \BibitemOpen
  \bibfield  {author} {\bibinfo {author} {\bibfnamefont {J.}~\bibnamefont
  {Koch}}, \bibinfo {author} {\bibfnamefont {T.~M.}\ \bibnamefont {Yu}},
  \bibinfo {author} {\bibfnamefont {J.}~\bibnamefont {Gambetta}}, \bibinfo
  {author} {\bibfnamefont {A.~A.}\ \bibnamefont {Houck}}, \bibinfo {author}
  {\bibfnamefont {D.~I.}\ \bibnamefont {Schuster}}, \bibinfo {author}
  {\bibfnamefont {J.}~\bibnamefont {Majer}}, \bibinfo {author} {\bibfnamefont
  {A.}~\bibnamefont {Blais}}, \bibinfo {author} {\bibfnamefont {M.~H.}\
  \bibnamefont {Devoret}}, \bibinfo {author} {\bibfnamefont {S.~M.}\
  \bibnamefont {Girvin}},\ and\ \bibinfo {author} {\bibfnamefont {R.~J.}\
  \bibnamefont {Schoelkopf}},\ }\bibfield  {title} {\bibinfo {title}
  {Charge-insensitive qubit design derived from the cooper pair box},\ }\href
  {https://doi.org/10.1103/PhysRevA.76.042319} {\bibfield  {journal} {\bibinfo
  {journal} {Phys. Rev. A}\ }\textbf {\bibinfo {volume} {76}},\ \bibinfo
  {pages} {042319} (\bibinfo {year} {2007})}\BibitemShut {NoStop}%
\bibitem [{\citenamefont {Rigetti}\ and\ \citenamefont
  {Devoret}(2010)}]{PhysRevB.81.134507}%
  \BibitemOpen
  \bibfield  {author} {\bibinfo {author} {\bibfnamefont {C.}~\bibnamefont
  {Rigetti}}\ and\ \bibinfo {author} {\bibfnamefont {M.}~\bibnamefont
  {Devoret}},\ }\bibfield  {title} {\bibinfo {title} {Fully microwave-tunable
  universal gates in superconducting qubits with linear couplings and fixed
  transition frequencies},\ }\href {https://doi.org/10.1103/PhysRevB.81.134507}
  {\bibfield  {journal} {\bibinfo  {journal} {Phys. Rev. B}\ }\textbf {\bibinfo
  {volume} {81}},\ \bibinfo {pages} {134507} (\bibinfo {year}
  {2010})}\BibitemShut {NoStop}%
\bibitem [{\citenamefont {Mundada}\ \emph {et~al.}(2019)\citenamefont
  {Mundada}, \citenamefont {Zhang}, \citenamefont {Hazard},\ and\ \citenamefont
  {Houck}}]{PhysRevApplied.12.054023}%
  \BibitemOpen
  \bibfield  {author} {\bibinfo {author} {\bibfnamefont {P.}~\bibnamefont
  {Mundada}}, \bibinfo {author} {\bibfnamefont {G.}~\bibnamefont {Zhang}},
  \bibinfo {author} {\bibfnamefont {T.}~\bibnamefont {Hazard}},\ and\ \bibinfo
  {author} {\bibfnamefont {A.}~\bibnamefont {Houck}},\ }\bibfield  {title}
  {\bibinfo {title} {Suppression of qubit crosstalk in a tunable coupling
  superconducting circuit},\ }\href
  {https://doi.org/10.1103/PhysRevApplied.12.054023} {\bibfield  {journal}
  {\bibinfo  {journal} {Phys. Rev. Appl.}\ }\textbf {\bibinfo {volume} {12}},\
  \bibinfo {pages} {054023} (\bibinfo {year} {2019})}\BibitemShut {NoStop}%
\bibitem [{\citenamefont {Kang}\ \emph {et~al.}(2024)\citenamefont {Kang},
  \citenamefont {Kim}, \citenamefont {Kim},\ and\ \citenamefont
  {Kwon}}]{Kang2024NewDO}%
  \BibitemOpen
  \bibfield  {author} {\bibinfo {author} {\bibfnamefont {J.}~\bibnamefont
  {Kang}}, \bibinfo {author} {\bibfnamefont {C.}~\bibnamefont {Kim}}, \bibinfo
  {author} {\bibfnamefont {Y.}~\bibnamefont {Kim}},\ and\ \bibinfo {author}
  {\bibfnamefont {Y.}~\bibnamefont {Kwon}},\ }\bibfield  {title} {\bibinfo
  {title} {New design of three-qubit system with three transmons and a single
  fixed-frequency resonator coupler},\ }\href
  {https://api.semanticscholar.org/CorpusID:274965131} {\bibfield  {journal}
  {\bibinfo  {journal} {Scientific Reports}\ }\textbf {\bibinfo {volume} {15}}
  (\bibinfo {year} {2024})}\BibitemShut {NoStop}%
\bibitem [{\citenamefont {Tripathi}\ \emph {et~al.}(2019)\citenamefont
  {Tripathi}, \citenamefont {Khezri},\ and\ \citenamefont
  {Korotkov}}]{PhysRevA.100.012301}%
  \BibitemOpen
  \bibfield  {author} {\bibinfo {author} {\bibfnamefont {V.}~\bibnamefont
  {Tripathi}}, \bibinfo {author} {\bibfnamefont {M.}~\bibnamefont {Khezri}},\
  and\ \bibinfo {author} {\bibfnamefont {A.~N.}\ \bibnamefont {Korotkov}},\
  }\bibfield  {title} {\bibinfo {title} {Operation and intrinsic error budget
  of a two-qubit cross-resonance gate},\ }\href
  {https://doi.org/10.1103/PhysRevA.100.012301} {\bibfield  {journal} {\bibinfo
   {journal} {Phys. Rev. A}\ }\textbf {\bibinfo {volume} {100}},\ \bibinfo
  {pages} {012301} (\bibinfo {year} {2019})}\BibitemShut {NoStop}%
\bibitem [{\citenamefont {Choi}\ \emph {et~al.}(2023)\citenamefont {Choi},
  \citenamefont {Abdelrahman}, \citenamefont {Slater}, \citenamefont {Parker},
  \citenamefont {Krantz}, \citenamefont {Sohn}, \citenamefont {Hassan},\ and\
  \citenamefont {Mueth}}]{Choi2023QuantumProAI}%
  \BibitemOpen
  \bibfield  {author} {\bibinfo {author} {\bibfnamefont {J.}~\bibnamefont
  {Choi}}, \bibinfo {author} {\bibfnamefont {M.~I.}\ \bibnamefont
  {Abdelrahman}}, \bibinfo {author} {\bibfnamefont {D.}~\bibnamefont {Slater}},
  \bibinfo {author} {\bibfnamefont {T.}~\bibnamefont {Parker}}, \bibinfo
  {author} {\bibfnamefont {P.}~\bibnamefont {Krantz}}, \bibinfo {author}
  {\bibfnamefont {P.}~\bibnamefont {Sohn}}, \bibinfo {author} {\bibfnamefont
  {M.~A.}\ \bibnamefont {Hassan}},\ and\ \bibinfo {author} {\bibfnamefont
  {C.}~\bibnamefont {Mueth}},\ }\bibfield  {title} {\bibinfo {title}
  {Quantumpro: An integrated workflow for the design of superconducting qubits
  using pathwave advanced design system (ads)},\ }\href
  {https://api.semanticscholar.org/CorpusID:265524908} {\bibfield  {journal}
  {\bibinfo  {journal} {2023 IEEE International Conference on Quantum Computing
  and Engineering (QCE)}\ }\textbf {\bibinfo {volume} {02}},\ \bibinfo {pages}
  {224} (\bibinfo {year} {2023})}\BibitemShut {NoStop}%
\bibitem [{\citenamefont {Lambert}\ \emph {et~al.}(2026)\citenamefont
  {Lambert}, \citenamefont {Giguère}, \citenamefont {Menczel}, \citenamefont
  {Li}, \citenamefont {Hopf}, \citenamefont {Suárez}, \citenamefont {Gali},
  \citenamefont {Lishman}, \citenamefont {Gadhvi}, \citenamefont {Agarwal},
  \citenamefont {Galicia}, \citenamefont {Shammah}, \citenamefont {Nation},
  \citenamefont {Johansson}, \citenamefont {Ahmed}, \citenamefont {Cross},
  \citenamefont {Pitchford},\ and\ \citenamefont {Nori}}]{LAMBERT20261}%
  \BibitemOpen
  \bibfield  {author} {\bibinfo {author} {\bibfnamefont {N.}~\bibnamefont
  {Lambert}}, \bibinfo {author} {\bibfnamefont {E.}~\bibnamefont {Giguère}},
  \bibinfo {author} {\bibfnamefont {P.}~\bibnamefont {Menczel}}, \bibinfo
  {author} {\bibfnamefont {B.}~\bibnamefont {Li}}, \bibinfo {author}
  {\bibfnamefont {P.}~\bibnamefont {Hopf}}, \bibinfo {author} {\bibfnamefont
  {G.}~\bibnamefont {Suárez}}, \bibinfo {author} {\bibfnamefont
  {M.}~\bibnamefont {Gali}}, \bibinfo {author} {\bibfnamefont {J.}~\bibnamefont
  {Lishman}}, \bibinfo {author} {\bibfnamefont {R.}~\bibnamefont {Gadhvi}},
  \bibinfo {author} {\bibfnamefont {R.}~\bibnamefont {Agarwal}}, \bibinfo
  {author} {\bibfnamefont {A.}~\bibnamefont {Galicia}}, \bibinfo {author}
  {\bibfnamefont {N.}~\bibnamefont {Shammah}}, \bibinfo {author} {\bibfnamefont
  {P.}~\bibnamefont {Nation}}, \bibinfo {author} {\bibfnamefont
  {J.}~\bibnamefont {Johansson}}, \bibinfo {author} {\bibfnamefont
  {S.}~\bibnamefont {Ahmed}}, \bibinfo {author} {\bibfnamefont
  {S.}~\bibnamefont {Cross}}, \bibinfo {author} {\bibfnamefont
  {A.}~\bibnamefont {Pitchford}},\ and\ \bibinfo {author} {\bibfnamefont
  {F.}~\bibnamefont {Nori}},\ }\bibfield  {title} {\bibinfo {title} {Qutip 5:
  The quantum toolbox in python},\ }\href
  {https://doi.org/https://doi.org/10.1016/j.physrep.2025.10.001} {\bibfield
  {journal} {\bibinfo  {journal} {Physics Reports}\ }\textbf {\bibinfo {volume}
  {1153}},\ \bibinfo {pages} {1} (\bibinfo {year} {2026})},\ \bibinfo {note}
  {quTiP 5: The Quantum Toolbox in Python}\BibitemShut {NoStop}%
\end{thebibliography}%

\end{document}